\documentclass[reprint,amsmath,amssymb,aps,showkeys]{revtex4-2}
\usepackage{graphicx}
\usepackage{physics}
\usepackage{xcolor}
\usepackage{mathptmx}
\usepackage[colorlinks=true
,urlcolor=blue
,citecolor=blue
,linkcolor=blue
]{hyperref}
\usepackage{float}
\newcommand{\etal}{\textit{et al.}}
\begin{document}
\preprint{APS/123-QED}
\title{Enhancing the force sensitivity of
squeezed light optomechanical interferometer}
\author{Sreeshna Subhash$^{1}$, Sanket Das$^{2}$, Tarak Nath Dey$^2$, Yong Li$^{3,4,5}$, Sankar Davuluri$^1$}%
\email{sankar@hyderabad.bits-pilani.ac.in}
\affiliation{%
${}^1$Department of Physics, Birla Institute of Technology and Science-Pilani, Hyderabad Campus, Hyderabad 500078, India\\
${}^2$Department of Physics, Indian Institute of Technology Guwahati, Guwahati 781039, Assam, India\\
${}^3$Beijing Computational Science Research Center, Beijing 100193, China\\
${}^4$Center for Theoretical Physics and School of Science, Hainan University, Haikou 570228, China\\
${}^5$Synergetic Innovation Center for Quantum Effects and Applications, Hunan Normal University, Changsha 410081, China}
\date{\today}
\begin{abstract}
Application of frequency-dependent squeezed vacuum improves the force sensitivity of optomechanical interferometer beyond the standard quantum limit by a factor of $e^{-r}$, where $r$ is the squeezing parameter. In this work, we show that the application of squeezed light along with quantum optical restoring force can enhance the sensitivity beyond the standard quantum limit by a factor of $\sqrt{e^{-2r}\zeta/4\Delta}$, where $0< \zeta/\Delta <1$, with $\zeta$ as the optomechanical cavity decay rate and $\Delta$ as the detuning between cavity eigenfrequency and driving field. The technique described in this article is restricted to frequencies much smaller than the resonance frequency of the optomechanical mirror.
\end{abstract}
\maketitle
\section{Introduction}
The quest to detect gravitational waves \cite{GIAZOTTO1989365,Yu2020} has revolutionized
precision measurements using an optical interferometer. The
laser interferometer gravitational wave detector is based on the coupling of optical modes with mechanical modes, which is known as optomechanics \cite{optomechanics1,RevModPhys.86.1391,Kippenberg:07,Barzanjeh2022}. With the miniaturization \cite{PhysRevLett.97.133601,miniaturizationOM3,cavityoptomechanics4nature} of mechanical mirrors, optomechanics has 
emerged as one of the best physical systems
to design ultra-precise sensors \cite{sensors,forcesensor,RevModPhys.68.755,RevModPhys.52.341}. Such a sensor can be designed by embedding optomechanical cavities into the arms of an optical interferometer \cite{PhysRevA.104.L031501,PhysRevX.10.031065}.
\par Shot noise and radiation pressure noise (RPN) \cite{LIGO2,PhysRevLett.45.75,Murch2008,Cripe2019}
are two major noises in optomechanics. Shot noise arises
from the randomness in the photon counting, while the RPN 
arises because of the randomness in the radiation pressure
force exerted on the mechanical mirror. Shot noise can be
decreased by increasing the laser power; however, this leads
to an increase in RPN. This trade-off between shot noise and
RPN imposes standard quantum limit (SQL) \cite{PhysRevA.94.013808,PhysRevA.34.3927}. Several techniques \cite{braginsky1980quantum,QND2,QND3,PhysRevLett.40.667,Clerk_2008,variationalmeasurement,PhysRevLett.105.123601,negativemassreferenceframe,doi:10.1126/science.1104149,doi:10.1126/sciadv.abg9204,doi:10.1126/science.1138007} were developed to overcome SQL.
One of the most popular methods is to use squeezed light \cite{PhysRevLett.117.030801,PhysRevLett.121.243601,Jaekel_1990,Ma2017,squeezing2,Lawrie2019,Lee2020,Aggarwal2020,Aasi2013,Safavi-Naeini2013,PhysRevLett.116.093602}. A
squeezed light \cite{Andersen_2016,squeezing3,squeezing4,squeezing5,Zhang2021,10.1103/PhysRevApplied.12.064024} is a special quantum state in which the uncertainty in one quadrature is decreased at the expense of
increased uncertainty in the other. Frequency-dependent squeezing \cite{PhysRevD.65.022002,PhysRevD.90.062006,Dutt:16} can improve the force sensitivity of optomechanical interferometer by a
factor \cite{Davuluri_2016} of $e^{-r}$ with $r$ being the squeezing parameter. In theory, $r$ can go up to infinity, but it is not so easy to obtain large $r$ experimentally. To our knowledge, the highest squeezing reported experimentally to date is $15$ dB \cite{PhysRevLett.117.110801}. In this article, we propose a method to improve the squeezed light optomechanical interferometer sensitivity by a factor of $\sqrt{e^{-2r}\zeta/4\Delta}$, here $\zeta$ is the cavity decay rate and $\Delta$ is detuning, for $0<\zeta/\Delta<1$. That means, for the same amount of squeezing or the same squeezing parameter $r$, the sensitivity is improved by a factor of $\sqrt{\zeta/4\Delta}$ beyond the best sensitivity achieved by squeezing alone. 
\section{Model}
\begin{figure}[hbt!]
\includegraphics[scale=.68]{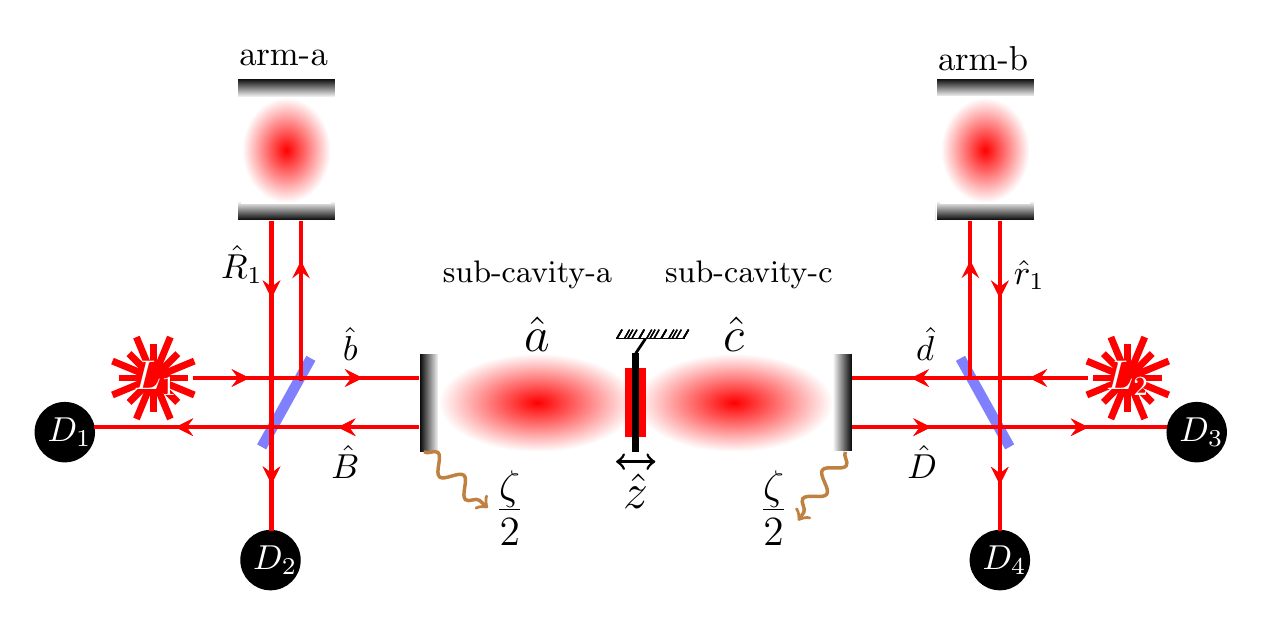}
\centering
\caption{Interferometer with a membrane in the middle. The optomechanical membrane is perfectly reflective.The optical fields in sub-cavity-a and sub-cavity-c are synthesized such that optical restoring force counters the fluctuations induced by radiation pressure force.
}
\label{fig1}
\end{figure}

Consider an optomechanical cavity with a perfectly reflective mechanical mirror in the middle \cite{Burgwal_2020} as shown in Fig. \ref{fig1}.
The mechanical mirror divides the total cavity into two sub-cavities, each with length $l$ and eigenfrequency $\omega_{e}$. The annihilation operators for optical fields inside the sub-cavities are given
by $\hat{a}$ and $\hat{c}$ as shown in Fig. \ref{fig1}. There is no tunneling of $\hat{a}$ into $\hat{c}$ and vice-versa as the mechanical mirror is perfectly
reflective. A co-sinusoidal classical force $f\cos(\omega_{f} t)$, with $\omega_{f}$ as frequency and $t$ as time, changes the position $\hat{z}$ of the mechanical mirror. The total Hamiltonian $\hat{H}$ of the  optomechanical cavity \cite{PhysRevA.51.2537} is given as
\begin{equation}\label{1}
\begin{split}
\hat{H}&=\frac{\hat{p}^{2}}{2m}+\frac{1}{2}m \omega_{m}^{2}\hat{z}^{2}+\hbar \omega_{e}\left(\hat{a}^{\dagger}\hat{a}+\frac{1}{2}\right)\left(1-\frac{\hat{z}}{l}\right)\\
&+\hbar \omega_{e} \left(\hat{c}^{\dagger}\hat{c}+\frac{1}{2}\right)\left(1+\frac{\hat{z}}{l}\right)-f\cos(\omega_{f} t)\hat{z}+\hat{H}_{r},
\end{split}
\end{equation}
where $\hat{p}$, $\omega_{m}$ and $m$ are momentum, eigenfrequency and the mass of the mechanical mirror, respectively. $\hat{H}_{r}$ is the Hamiltonian for the environment and its coupling with the optomechanical cavity. $\hbar$ is the reduced Planck's constant. The optical fields $\hat{a}$ and $\hat{c}$ are driven by input fields with annihilation operators $\hat{b}$ and $\hat{d}$, respectively. The dynamics of the optomechanical interaction are given as
\begin{eqnarray}
\label{eqn:abar}
\dot{\hat{a}}=\left(-i\Delta+ig\hat{z}-\frac{\zeta}{2}\right)\hat{a}+\sqrt{\zeta} \hat{b},
\end{eqnarray}
\begin{eqnarray}
\label{eqn:cbar}
\dot{\hat{c}}=\left(-i\Delta-ig\hat{z}-\frac{\zeta}{2}\right)\hat{c}+\sqrt{\zeta} \hat{d},
\end{eqnarray}
\begin{eqnarray}
\label{eqn:zbar}
m\left(\ddot{\hat{z}}+\gamma \dot{\hat{z}}+\omega_{m}^{2}\hat{z}\right)=\hbar g \left(\hat{a}^{\dagger}\hat{a}-\hat{c}^{\dagger}\hat{c}\right)+\hat{\varpi}+f\cos(\omega_{f}t),
\end{eqnarray}
\begin{eqnarray}\label{5}
\hat{B}=\hat{b}-\sqrt{\zeta}\hat{a}, \;\;\;\;\;\;  \hat{D}=\hat{d}-\sqrt{\zeta}\hat{c},
\end{eqnarray}
where $g=\omega_{e} /l$, same for both the sub-cavities \cite{PhysRevLett.109.063601,PhysRevLett.104.083901}, $\zeta$ is the cavity decay rate, $\gamma$ is the decay rate
of mechanical mirror, $\hat{\varpi}$ is the thermal noise operator for the mechanical mirror, and $\Delta = \omega_{e} - \omega_{d}$, with $\omega_{d}$ as the frequency of the external driving fields $\hat{b}$ and $\hat{d}$. The Eq. (\ref{5}) comes from the input-output relations. The operators $\hat{b}$ and $\hat{d}$ are normalized such that  their optical powers are given by $\hbar \omega_{d}\langle\hat{b}^{\dagger}\hat{b}\rangle$ and $\hbar \omega_{d}\langle\hat{d}^{\dagger}\hat{d}\rangle$, respectively. The operators $\hat{B}$ and $\hat{D}$ are the annihilation operators for the output field from sub-cavity-a and sub-cavity-b, respectively. We follow the standard procedure of linearizing Eq. (\ref{eqn:abar}) to Eq. (\ref{5}) by writing the operator $\hat{O}=\bar{O}+\hat{\delta}_{O}$ ($O=a,b,c,d,B,D$) with $\bar{O}$ represents the steady state while $\hat{\delta}_{O}$ represents the fluctuation. By using this notation, Eq. (\ref{eqn:abar}) to Eq. (\ref{5}) can be solved to obtain 
\begin{equation}\label{6}
\bar{a}=\frac{\sqrt{\zeta }\bar{b}}{i\Delta-ig\bar{z}+\frac{\zeta}{2}};
\bar{c}=\frac{\sqrt{\zeta} \bar{d}}{i\Delta+ig\bar{z}+\frac{\zeta}{2}};
\bar{z}=\frac{\hbar g \left(|\bar{a}|^{2}-|\bar{c}|^{2}\right)}{m\omega_{m}^{2}}.
\end{equation}
There is no $f\cos(\omega_{f}t)$ term in Eq. (\ref{6}) as it is treated like a small perturbation and included in the fluctuations. The $\bar{z}$ in $\bar{a}$ and $\bar{c}$ leads to optomechanical bi-stability \cite{PhysRevLett.51.1550}. In Eq. (\ref{6}) bi-stability can be avoided  by choosing $|\bar{a}|^{2}=|\bar{c}|^{2}$. Then the average radiation pressure force on the mechanical mirror from both the sub-cavities is equal  but opposite in direction. Hence, $\bar{z}$ becomes zero. We assume the beam-splitters in Fig. \ref{fig1} are 50:50. The input fields $\hat{b}$ and $\hat{d}$ are phase adjusted such that, $\hat{b}=\left(\hat{E}+i\hat{V}\right)e^{-i\phi}/\sqrt{2}$ and $\hat{d}=\left(\hat{F}+i\hat{U}\right)e^{-i\phi}/\sqrt{2}$, where $\phi=\tan^{-1}\left(-2\Delta/\zeta\right)$, $\hat{E}$ and $\hat{F}$ are the laser field annihilation operators while $\hat{V}$ and $\hat{U}$ are the vacuum field annihilation operators. Then the steady state cavity fields $\bar{a}=\sqrt{\zeta}\bar{E}/\sqrt{2\left(\Delta^{2}+\zeta^{2}/{4}\right)}$ and $\bar{c}=\sqrt{\zeta}\bar{F}/{\sqrt{2\left(\Delta^{2}+\zeta^{2}/{4}\right)}}$ can be set to be real by taking $\bar{E}$ and $\bar{F}$ as real where $\bar{E}$ and $\bar{F}$ are the mean values of $\hat{E}$ and $\hat{F}$, respectively. By using Eq. (\ref{5}) and Eq. (\ref{6}), we can write
\begin{equation}\label{7}
\bar{B}=\bar{D}=-\frac{\bar{E}}{\sqrt{2}}e^{i\phi}.
\end{equation}
 As there are no external losses in the sub-cavities, Eq. (\ref{7})  imply that the average optical power of the output field and input field are same. As $\bar{z}=0$, Eq. (\ref{7}) is not influenced by optomechanical interaction but the phase $\phi$ appears because of the input laser detuning from the cavity resonance. The equations of motion for fluctuations are given as
\begin{equation}\label{eqm1}
\dot{\hat{\delta}}_{a}=\left(-i\Delta-\frac{\zeta}{2}\right)\hat{\delta}_{a}+ig\bar{a}\hat{\delta}_{z}+\sqrt{\zeta}\hat{\delta}_{b},
\end{equation}
\begin{equation}\label{eqm2}
\dot{\hat{\delta}}_{c}=\left(-i\Delta-\frac{\zeta}{2}\right)\hat{\delta}_{c}-ig\bar{a}\hat{\delta}_{z}+\sqrt{\zeta}\hat{\delta}_{d},
\end{equation}
\begin{equation}\label{eqm3}
\begin{split}
\ddot{\hat{\delta}}_{z}+\gamma \dot{\delta}_{z}+\omega_{m}^{2}\hat{\delta}_{z}=&\frac{\hbar g \bar{a}}{m}\left(\hat{\delta}_{a}^{\dagger}+\hat{\delta}_{a}-\hat{\delta}_{c}^{\dagger}-\hat{\delta}_{c}\right)+
\frac{\hat{\varpi}}{m}\\
&+\frac{f}{m}\cos(\omega_{f}t).
\end{split}
\end{equation}
Note that we used the relation $\bar{a}=\bar{a}^{*}=\bar{c}$ in writing Eq. (\ref{eqm1}) to Eq. (\ref{eqm3}). The super fix symbols $``*"$ and $``\dagger"$ are  represents complex conjugate and adjoint operations, respectively. It is useful to write Eq. (\ref{eqm1}) to Eq. (\ref{eqm3}) in a simplified version as
\begin{equation}\label{8}
\dot{\hat{M}}=\left(-i\Delta-\frac{\zeta}{2}\right)\hat{M}+i2g\bar{a}\hat{\delta}_{z}+\sqrt{\zeta}\hat{M}_{1},
\end{equation}
\begin{equation}\label{9}
m\left(\ddot{\hat{\delta}}_{z}+\gamma \dot{\hat{\delta}}_{z}+\omega_{m}^{2}\hat{\delta}_{z}\right)=\hbar g \bar{a}\left(\hat{M}+\hat{M}^{\dagger}\right)+\hat{\varpi}+f\cos\left(\omega_{f}t\right),
\end{equation}
where $\hat{M}=\hat{\delta}_{a}-\hat{\delta}_{c}$, $\hat{M}_{1}=\hat{\delta}_{b}-\hat{\delta}_{d}$. Position of the mechanical mirror can be inferred by measuring the phase of the output field at the detectors $D_{1}$ and $D_{2}$ or $D_{3}$ and $D_{4}$. However, as we are dealing with $\hat{M}$, which is a joint operator of $\hat{\delta}_{a}$ and $\hat{\delta}_{c}$, we measure the relative phase between $\hat{B}$ and $\hat{D}$. Hence the general homodyne measurement is slightly modified to measure $\hat{Q}$, which is given as 
\begin{equation}\label{10}
\hat{Q}=\left(\hat{B}^{\dagger}\hat{R}_{1}+\hat{R}_{1}^{\dagger}\hat{B}\right)-\left(\hat{D}^{\dagger}\hat{r}_{1}+\hat{r}_{1}^{\dagger}\hat{D}\right),
\end{equation}
where $\hat{R}_{1}$ and $\hat{r}_{1}$ are reference fields at the output of the optical cavities in arm-a and arm-b, respectively. These optical cavities are on resonance with the incoming fields and have rigidly fixed mirrors with the upper mirrors being perfectly reflective while the lower mirrors have the decay rate $\zeta$. The reference fields can be written in terms of input fields as

\begin{equation}\label{11}
\hat{R}_{1}(\omega)=H\frac{i\hat{E}(\omega)+\hat{V}(\omega)}{\sqrt{2}},\;\; \hat{r}_{1}(\omega)=H\frac{i\hat{F}(\omega)+\hat{U}(\omega)}{\sqrt{2}},
\end{equation}
where $H={\left(i\omega+\zeta/2\right)}/{\left(i\omega-\zeta/2\right)}$, with $\omega$ as Fourier frequency. Using Eq. (\ref{8}), Eq. (\ref{9}) and Eq. (\ref{5}), after some mathematical manipulation, the quantum fluctuation in the output fields is given as
\begin{equation}\label{12}
\begin{split}
\hat{Y}_{B}(\omega)-\hat{Y}_{D}(\omega)&=G_{1}\left(\hat{\delta}_{b}^{\dagger}(-\omega)-\hat{\delta}_{d}^{\dagger}(-\omega)\right)+\\
&G_{2}\left(\hat{\delta}_{b}(\omega)-\hat{\delta}_{d}(\omega)\right)+G_{3}\hat{\varpi}(\omega),
\end{split}
\end{equation}
where $\hat{Y}_{O}(\omega)=i\left[\hat{\delta}_{O}^{\dagger}(-\omega)-\hat{\delta}_{O}(\omega)\right],$ with $O=B,D $ and
\begin{equation}
\begin{split}
G_{1}=i&+\frac{i\zeta-\frac{(\alpha-\Delta)\zeta}{i\omega-{\zeta}/{2}}}{\left(i\omega-\frac{\zeta}{2}-\frac{(\alpha-\Delta)\Delta}{i\omega-{\zeta}/{2}}\right)}, G_{3} = \frac{\sqrt{\zeta}\frac{4g\bar{a}}{m(\omega_{m}^{2}-\omega^{2}-i\gamma\omega)}}{\left(i\omega-\frac{\zeta}{2}-\frac{(\alpha-\Delta)\Delta}{i\omega-{\zeta}/{2}}\right)},\nonumber\\
&\;\;\;\;\;\;\;\;G_{2}=-i+\frac{-i\zeta -\frac{(\alpha-\Delta)\zeta}{i\omega-{\zeta}/{2}}}{\left(i\omega-\frac{\zeta}{2}-\frac{(\alpha-\Delta)\Delta}{i\omega-{\zeta}/{2}}\right)},\nonumber
\end{split}
\end{equation}
with $\alpha=4\hbar g^{2}\bar{a}^{2}/[m(\omega_{m}^{2}-\omega^{2}-i\gamma \omega)]$. Substituting Eq. (\ref{12}) in the quantum fluctuation $\hat{\delta}_{Q}$ part of Eq. (\ref{10}) gives
\begin{equation}\label{13}
\begin{split}
\hat{\delta}_{Q}(\omega)&= \frac{\bar{E}}{\sqrt{2}}\left[\hat{Y}_{B}(\omega)-\hat{Y}_{D}(\omega)\right]+\bar{B}^{*}\left[\hat{\delta}_{R_{1}}(\omega)-\hat{\delta}_{r_{1}}(\omega)\right]\\
&+\bar{B}\left[\hat{\delta}_{R_{1}}^{\dagger}(-\omega)-\hat{\delta}_{r_{1}}^{\dagger}(-\omega)\right].
\end{split}
\end{equation}
We have used the relation $\bar{B}= \bar{D}$ in writing Eq. (\ref{13}). The fluctuations $\hat{\delta}_{R_{1}}$ and $\hat{\delta}_{r_{1}}$ in the reference fields are given as $\hat{\delta}_{R_{1}}(\omega)= H[i\hat{\delta}_{E}(\omega)+\hat{\delta}_{V}(\omega)]/\sqrt{2}$, $\hat{\delta}_{r_{1}}(\omega)=H[i\hat{\delta}_{F}(\omega)+\hat{\delta}_{U}(\omega)]/\sqrt{2}$. The cavities in arm-a and arm-b have rigidly fixed mirrors and they do not have any external losses. Hence the steady state reference fields are given as $\bar{R}_{1}=\bar{r}_{1}=i\bar{E}/\sqrt{2}$ (because $\bar{E}=\bar{F}$). The noise spectral density $S_{QQ}$ is given by Eq. (\ref{13}) according to the relation $\langle[\hat{\delta}_{Q}(\omega)]^{\dagger}\hat{\delta}_{Q}(\omega_{1})\rangle=S_{QQ}(\omega)\delta(\omega+\omega_{1})$.

 The action of $f \cos(\omega_{f} t)$ changes the equilibrium position of the mechanical mirror leading to signal $\bar{Q}$ as
\begin{equation}\label{14}
\bar{Q} = \frac{\bar{E}f}{2\sqrt{2}}\left[G_{3}(-\omega_{f})e^{i\omega_{f}t}+G_{3}(\omega_{f})e^{-i\omega_{f}t}\right].
\end{equation}
As the classical force $f\cos(\omega_{f}t)$ drives the mechanical mirror at frequency $\omega_{f}$, the $\bar{Q}$ is also oscillating at the same
frequency. Hence the force sensitivity $F_{s}$ at $\omega_{f}$ is given as
\begin{equation}\label{Eq.F_s}
F_{s}=\frac{\sqrt{S_{QQ}(-\omega_{f})+S_{QQ}(\omega_{f})}}{\bar{E}|G_{3}(\omega_{f})|/\sqrt{2}}.
\end{equation}
\section{RESULTS}
As a first step, we establish different varieties of noises and how they influence the force sensitivity given in Eq. (\ref{Eq.F_s}). For this, we set $\Delta=0$, $\omega=\omega_{f}$ in Eq. (\ref{Eq.F_s}) and estimate the force sensitivity $F_{o}$ as 
\begin{equation}\label{27}
F_{o}=\frac{m\omega_{m}^{2}\zeta}{4g}\sqrt{\frac{1}{2|\bar{E}|^{2}}+\frac{256\hbar^{2}g^{4}}{m^{2}\omega_{m}^{4}\zeta^{4}}|\bar{E}|^{2}+\frac{16 \hbar g^{2} \omega_{f}\gamma}{m\omega_{m}^{4}\zeta^{2}}}.
\end{equation}
The first term on the right-hand side (RHS) of Eq. (\ref{27}) gives the shot noise, while the second and third term gives the RPN and thermal noise, respectively. Temperature is assumed to be zero Kelvin in Eq. (\ref{27}). The contribution from shot noise and RPN compete in Eq. (\ref{27}) leading to SQL at an input intensity $I_{opt}$. Using Eq. (\ref{27}), the $I_{opt}$ can be estimated as
\begin{equation}\label{28}
I_{opt}=\frac{m\omega_{m}^{2}\zeta^{2}}{16{\sqrt{2}}\hbar g^{2}}.
\end{equation}
A prominent property of Eq. (\ref{27}) is its dependence on $|\bar{E}|^{2}$. For $|\bar{E}|^{2}>I_{opt}$ the shot noise contribution decreases but the RPN increases, similarly for $|\bar{E}|^{2}<I_{opt}$ the RPN decreases but the shot noise increases. Hence in Eq. (\ref{27}),
for best sensitivity, we must set $|\bar{E}|^{2}=I_{opt}$ which enforces SQL. Substituting  Eq. (\ref{28}) into Eq. (\ref{27}) gives the force sensitivity at $\Delta=0$ as $F_{1}=\sqrt[4]{2}F_{sql}$, where $F_{sql}=\sqrt{\hbar m \omega_{m}^{2}}$.
\par Equation (\ref{27}) establishes the presence of shot noise, RPN, and thermal noise. The objective of this manuscript is not only to go beyond the SQL but also to break the squeezed light limit. As a first step we describe using optical restoring force to suppress RPN when laser and vacuum are input fields \cite{davuluri2017quantum}. After that, we apply squeezed vacuum technique in conjunction
with optical restoring force to go beyond the squeezed light limits. As $\alpha$ is the only
variable with optomechanical coupling $g$ in Eq. (\ref{12}), any contribution to RPN must come from $\alpha$. By setting $\Delta - \alpha=0$, the contribution to the RPN from the real part of $\alpha$ is eliminated. However, $\Delta$ is a real quantity while $\alpha$ is complex.
\begin{equation}\label{16}
\alpha=\frac{4\hbar g^{2}\bar{a}^{2}e^{i\tan^{-1}\epsilon}}{m\sqrt{\left(\omega_{m}^{2}-\omega^{2}\right)^{2}+\gamma^{2}\omega^{2}}}\approx \frac{4\hbar g^{2}\bar{a}^{2}}{m\omega_{m}^{2}}(1+i\epsilon),
\end{equation}
where $\epsilon=\gamma \omega/\left(\omega_{m}^{2}-\omega^{2}\right)$ and the last term in Eq. (\ref{16}) is valid for frequencies (we refer this as low frequency regime) much smaller than $\omega_{m}$. Hence it is impossible to achieve $\Delta-\alpha=0$. However, the strength of the RPN can be significantly reduced by setting $\Delta-\mathcal{R}(\alpha)=0$ for $\omega \ll \omega_{m}$ and $\gamma \ll \omega_{m}$, where $\mathcal{R}$ stands for the real part. Setting $\Delta-\mathcal{R}(\alpha)=0$ eliminates the RPN contribution from $\mathcal{R}(\alpha)$. The residual RPN from the imaginary part of $\alpha$ is significantly less than the shot noise in the low frequency regime as $\gamma\omega\ll \omega_{m}^{2}$.  As a result, the method described in this manuscript is strictly limited to
frequencies much smaller than the resonance frequency of the
mechanical mirror. At these lower frequencies, the force
sensitivity is less because of large RPN. Hence suppressing RPN in the low frequencies is quite important. Note that $\bar{Q}$ is oscillating at $\omega_{f}$, so we only need to bother about noise at $\omega_{f}$. As $\omega_{f} \ll \omega_{m}$, we are interested in finding noise where $\epsilon\ll 1$ is already satisfied. Hence setting $\Delta=\mathcal{R}(\alpha)$ should suppress RPN in our system. 
The force sensitivity $F_{s}$ is given as
\begin{equation}\label{19}
F_{s}=\frac{m\omega_{m}^{2}\sqrt{\Delta^{2}+{\zeta^{2}}/{4}}}{2g} \sqrt{\frac{1}{2|\bar{E}|^{2}}+\frac{4g^{2}\hbar \omega_{f}\gamma\coth\left(\frac{\hbar \omega_{f}}{2k_{B}T}\right)}{m\omega_{m}^{4}\left(\Delta^{2}+{\zeta^{2}}/{4}\right)}}.
\end{equation}
There is no RPN in Eq. (\ref{19}) as it is suppressed. We use the thermal correlation  \cite{PhysRevA.63.023812} $\left\langle {\hat\varpi}(\omega){\hat\varpi}(\omega') \right\rangle=\hbar m \omega \gamma[1+\coth(\hbar \omega/2k_{B}T)]\delta(\omega+\omega')$ with temperature $T$, and $k_{B}$ as Boltzmann constant. We simplified Eq. (\ref{19}) by assuming that $1>\zeta/\Delta > \zeta^{2}/\Delta^{2} \gg \gamma\omega_{f}/\omega_{m}^{2}$. The condition that $\zeta/\Delta$ should lie between 1 and $\epsilon$ is not necessary for RPN suppression but required for improving $F_{s}$ beyond SQL. The first term of the RHS of Eq. (\ref{19}) gives shot noise contribution while the second term gives the thermal noise contribution. The shot noise in Eq. (\ref{19}) can be decreased by increasing the intensity, however, the input intensity is constrained by the condition $\Delta=\mathcal{R}(\alpha)$ as
\begin{equation}\label{20}
\Delta=\frac{4\hbar g^{2}|\bar{a}|^{2}}{m\omega_{m}^{2}}\;\implies \;2|\bar{E}|^{2}=\frac{m\omega_{m}^{2}\left(\Delta^{2}+{\zeta^{2}}/{4}\right)\Delta}{\hbar g^{2} \zeta}.
\end{equation}
Substituting Eq. (\ref{20}) into Eq. (\ref{19}) gives the best force sensitivity $F_{2}$ achievable as
\begin{equation}\label{21}
F_{2}=F_{sql}\sqrt{\frac{\zeta}{4\Delta}+\frac{\gamma\omega_{f}}{\omega_{m}^{2}}}.
\end{equation}
We assumed $T=0K$ in Eq. (\ref{21}). Note that as $\gamma\omega_{f}/\omega_{m}^{2}\ll\zeta/4\Delta$, $F_{s}$ is better than $F_{sql}$ by a factor of $\sqrt{\zeta/4\Delta}$. The intensity in Eq. (\ref{20}) is larger than $I_{opt}$ by a factor of $(4\Delta^{2}/\zeta^{2}+1)2\sqrt{2}\Delta/\zeta$. With suppression of RPN, we are able to increase the intensity beyond $I_{opt}$. However the signal in Eq. (\ref{14}) is reduced by a factor of $1/\sqrt{4\Delta^{2}/\zeta^{2}+1}$. Combining these two factors, we observe an improvement by a factor of $\sqrt{\zeta/4\Delta}$ beyond $F_{sql}$. 
\par
\begin{figure}[hbt!]
\centering\
\includegraphics[scale=0.65]{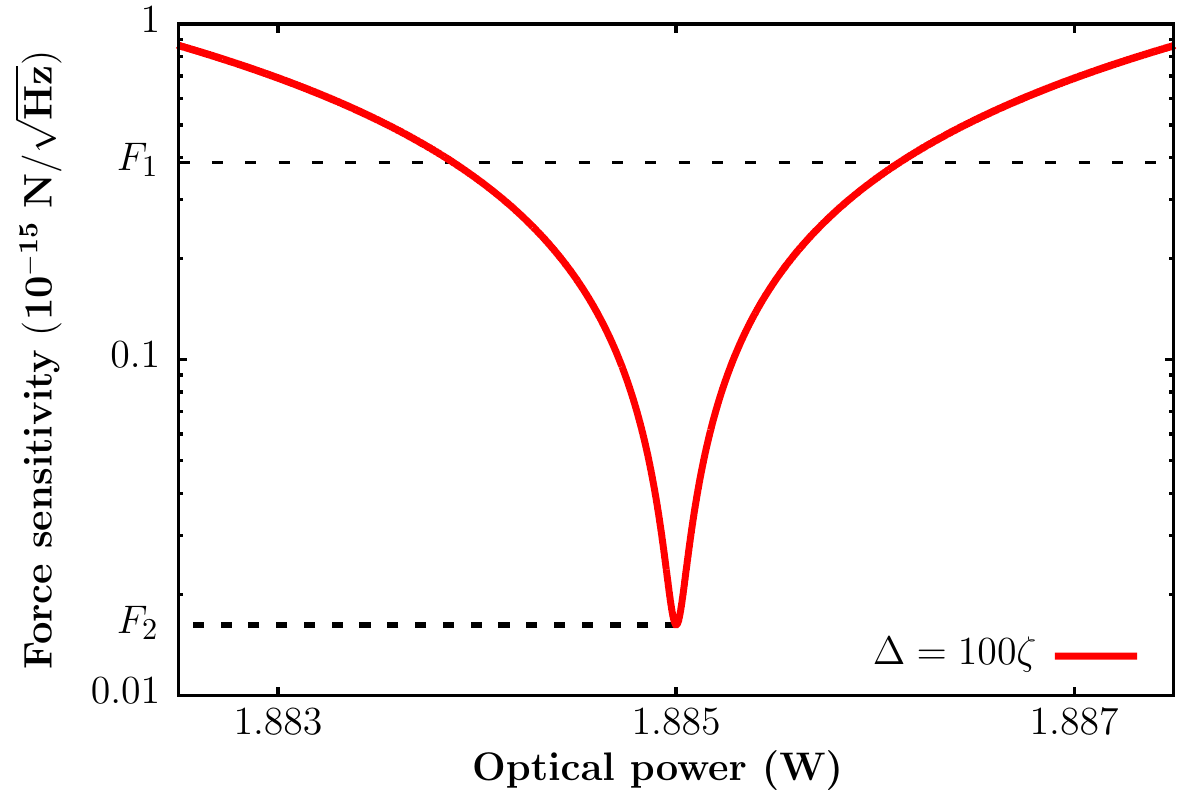}
\caption{Variation of force sensitivity as a function of optical power. RPN is suppressed by setting $\mathcal{R}(\alpha)\simeq\Delta$. The $F_{2}$ represents the force sensitivity improved by a factor of $\sqrt{\zeta/4\Delta}$ over $F_{sql}$.  The lowest point of the curve gives the best force sensitivity $F_{2}$ ($1.62\times10^{-17}$N/$\sqrt{\mbox{Hz}}$) at optical power $1.885$ W.} \
\label{dip}
\end{figure}

 A plot of Eq. (\ref{Eq.F_s}) as a function of input laser power is shown in Fig. \ref{dip}. In the plot, parameters are chosen such that the condition in Eq. (\ref{16}) is satisfied, $1>\zeta^{2}/\Delta^{2}\gg\gamma\omega_{f}/\omega_{m}^{2}$ and $\mathcal{R}(\alpha)\simeq\Delta$, so the RPN is suppressed. Hence the best force sensitivity $F_{2}$ in Fig. \ref{dip} is improved beyond $F_{1}$. 
\par The simulation parameters for Fig. \ref{dip} are : $m=10^{-7}$Kg, $\omega_{m}=10^{5}$Hz, $\omega_{f}=100$Hz, $\omega_{d}=6\pi/5\times10^{15}$Hz, $\Delta=100\zeta$, $\zeta=10^{6}$Hz,
$\gamma=1$Hz, $g=10^{18}$Hz/m. The optical power corresponding to the lowest point in Fig. \ref{dip} is given by $\hbar \omega_{d}|\bar{E}|^{2}$, where $|\bar{E}|^{2}$ is given by Eq. (\ref{20}). The best force sensitivity in Fig. \ref{dip} is exactly equal to $F_{2}$ which is given in Eq. (\ref{21}).
\subsection{Squeezing}
\par A squeezed light is a non-classical state \cite{doi:https://doi.org/10.1002/9781119009719.ch5,doi:10.1063/5.0024341} which allows shrinking uncertainty in one quadrature at the expense of increased uncertainty in the corresponding conjugate quadrature. On the other hand, the competitive behaviour between shot noise and RPN in optomechanics arises because of interplay between canonically conjugate quadratures \cite{PhysRevD.1.3217,PhysRevD.4.1925}. Hence
by eliminating RPN from Eq. (\ref{13}), we eliminated the interplay between the canonically conjugate quadratures in our system. 
 This allows us to use squeezed states to further enhance the force sensitivity without worrying about the increased
uncertainty from its conjugate quadrature. The most interesting aspect, as shown below, is that the overall force sensitivity
is better than what squeezed light alone can achieve.
\par The force sensitivity in Eq. (\ref{21}) is derived by assuming that the input fields are vacuum and laser fields. Now lets
squeeze the vacuum field \cite{Otterpohl:19,Ast:13,Aoki:06,PhysRevLett.124.193601} entering through the empty port of the interferometer so that
\begin{equation}\label{22}
\ket{U}_\xi=e^{\frac{1}{2}\left(\xi^{*}\hat{U}\hat{U}-\xi\hat{U}^{\dagger}\hat{U}^{\dagger}\right)}\ket{0},\;
\ket{V}_\xi=e^{\frac{1}{2}\left(\xi^{*}\hat{V}\hat{V}-\xi\hat{V}^{\dagger}\hat{V}^{\dagger}\right)}\ket{0},
\end{equation}
where $\xi=r e^{i\theta}$ with $r$ as the squeezing parameter and $\theta$ as the squeezing angle. Using Eq. (\ref{22}) and Eq. (\ref{13}), the symmetrized noise power spectral density $N_{\xi}$ of shot noise and RPN is given as
\begin{equation}\label{23}
N_{\xi}=2|\bar{E}|^{2}\left[\cosh(2r)-\sinh(2r)\cos(\theta-2\phi)\right]=2|\bar{E}|^{2}e^{-2r}.
\end{equation}
The final result in Eq. (\ref{23}) is obtained by considering frequency-dependent squeezing such that the squeezing angle $\theta=2\phi$. As the squeezing is implemented only on the vacuum field, which makes no contribution to $\bar{Q}$, the signal remains
same as given in Eq. (\ref{14}). Hence, with the squeezed vacuum,
the force sensitivity $F_{\xi}$
is given as
\begin{equation}\label{24}
\noindent
F_{\xi}=\frac{m\omega_{m}^{2}\sqrt{\Delta^{2}+\zeta^{2}/4}}{2g} \sqrt{\frac{e^{-2r}}{2|\bar{E}|^{2}}+\frac{4g^{2}\hbar \gamma \omega_{f}}{m\omega_{m}^{4}\left(\Delta^{2}+\zeta^{2}/4\right)}}.
\end{equation}
Substituting Eq. (\ref{20}) into Eq. (\ref{24}) gives the force sensitivity $F_{3}$ as
\begin{equation}\label{25}
F_{3} = F_{sql}\sqrt{\frac{\zeta}{4\Delta}e^{-2r}+\frac{\gamma\omega_{f}}{\omega_{m}^{2}}}.
\end{equation}
The RHS of Eq. (\ref{25}) shows that the sensitivity is improved by a factor of $\sqrt{e^{-2r}\zeta/4\Delta}$ beyond $F_{sql}$. In Eq. (\ref{25}), the squeezed light leads to $e^{-r}$ improvement while the optical restoring force leads to $\sqrt{\zeta/4\Delta}$ improvement. Hence using squeezed light in combination with quantum optical restoring force can enhance the interferometer performance beyond the squeezed light limit by a factor of $\sqrt{\zeta/4\Delta}$. It is worth mentioning that the biggest challenge with using squeezed states to improve optical interferometer is generating squeezed states with large $r$ value. To our knowledge, experimentally, the highest squeezing realized so far is $15$ dB. In this scenario Eq. (\ref{25}) presents an alternate approach to improve the squeezed light interferometer sensitivity not only by increasing $r$ but also by minimizing $\zeta/\Delta$ factor.
This point can be further illustrated by rewriting Eq. (\ref{25}) as
\begin{equation}\label{30ref}
F_{3}=F_{sql}\sqrt{e^{-2r_{eff}}+\frac{\gamma \omega_{f}}{\omega_{m}^{2}}},
\end{equation}
where $r_{eff}=r+\mbox{ln}\left(4\Delta/\zeta\right)/2$ is the effective squeezing parameter.  Hence for an input squeezing of $r$, the sensitivity of
the optomechanical interferometer is improved by an effective
squeezing of $r_{eff}$. At the same time, it must be noted that this method is strictly limited to frequencies much smaller than the resonance frequency of the mechanical mirror.
\par
\begin{figure}[hbt!]
\includegraphics[scale=0.65]{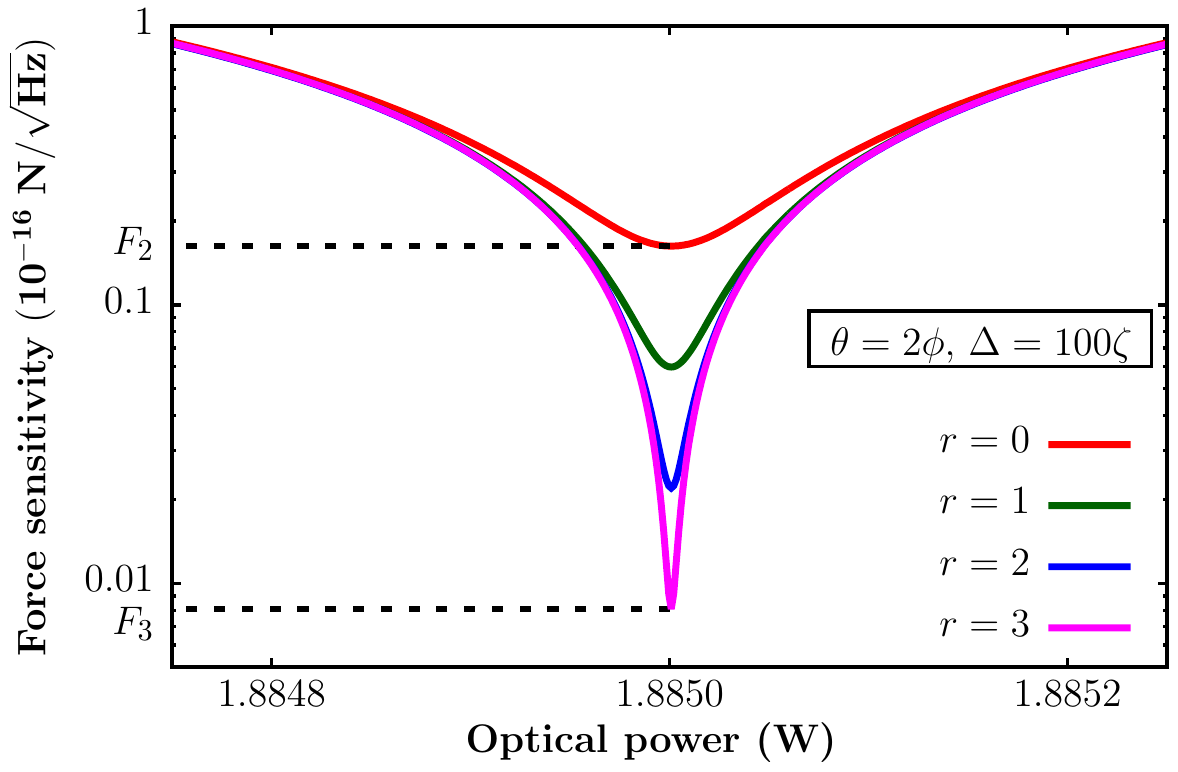}
\caption{Variation of force sensitivity as a function of optical power with different squeezing parameter $r$. The squeezing angle $\theta$ is fixed at $2\phi$. RPN is suppressed by setting $\mathcal{R}(\alpha)\simeq\Delta$. The lowest point of each curve is given by $F_{3}$. For $r=0$ curve, $F_{3}$ is same as $F_{2}$.}
\label{diff_r}
\end{figure}

\begin{figure}[hbt!]
\includegraphics[scale=0.65]{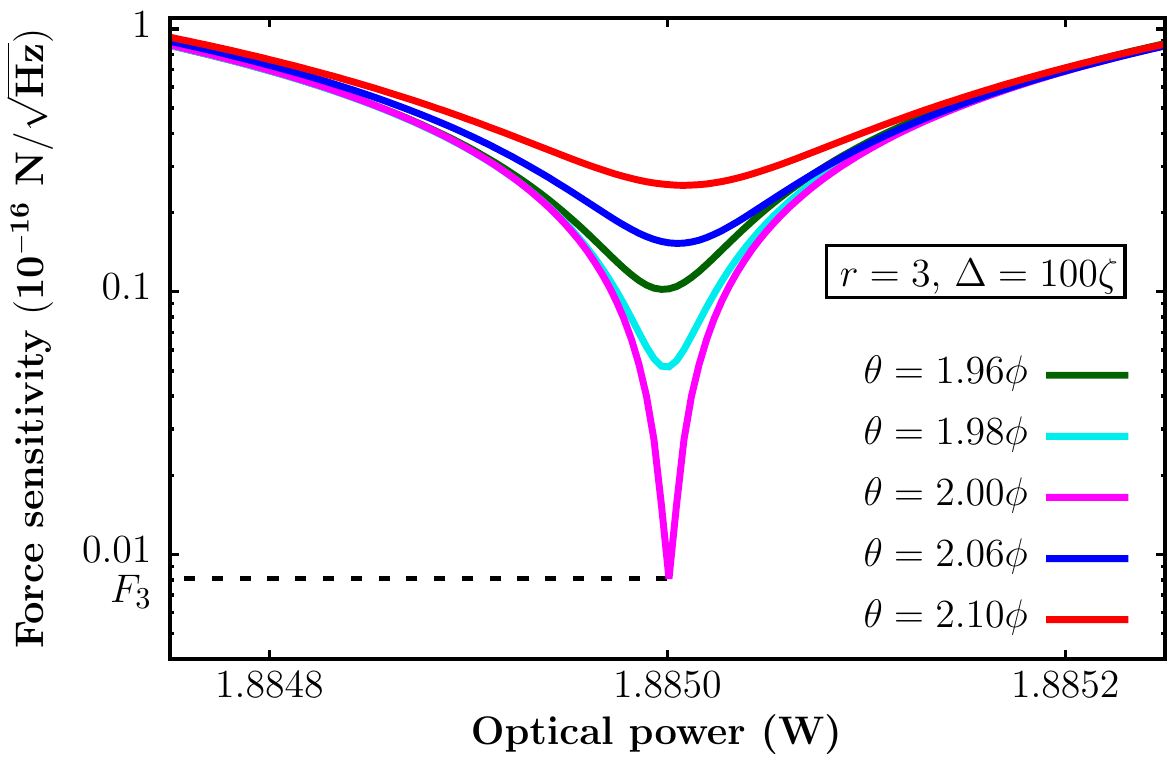}
\caption{Variation of force sensitivity as a function of optical power with different squeezing angles $\theta$. Squeezing parameter $r$ is fixed to $3$. RPN is suppressed by setting $\mathcal{R}(\alpha)\simeq\Delta$. Best sensitivity is achieved when $\theta=2\phi$.}
\label{diff_t}
\end{figure}

 The Eq. (\ref{25}), Eq. (\ref{23}),
Eq. (\ref{21}), and Eq. (\ref{20}) are analytically simplified results from Eq. (\ref{Eq.F_s}). These equations together illustrate the  improved sensitivity in the squeezed light optomechanical interferometer. To obtain a broader understanding on variation of sensitivity under various conditions, we plot Eq. (\ref{Eq.F_s}) as a function of various parameters such as input power, $r$, and $\theta$ in Fig. \ref{diff_r} and Fig. \ref{diff_t}. The results and conclusion from the plots agree with the simplified
results (Eq. (\ref{25}), Eq. (\ref{23}),
Eq. (\ref{21}), and Eq. (\ref{20})). Fig. \ref{diff_r} and Fig. \ref{diff_t} shows the enhanced force sensitivity beyond squeezed light limit at two different cases.\\
\\
Case (\romannumeral 1) : Figure. \ref{diff_r} shows the plot of Eq. (\ref{Eq.F_s})  when the input vacuum is squeezed for different $r$ values $i.e.,$ $r=0,1,2,3$. The plotting parameters chosen such that $1>\zeta^{2}/\Delta^{2}\gg\gamma\omega_{f}/\omega_{m}^{2}$, Eq. (\ref{16}) is satisfied and $\mathcal{R}(\alpha)\simeq\Delta$. So that RPN is suppressed. The squeezing angle is fixed at $\theta=2\phi$ and the squeezing parameter $r$ is varied for each plot. Increasing $r$ value increases the overall force sensitivity. The best sensitivity in each curve corresponds to $F_{3}$ which is given by Eq. (\ref{25}). The optical power $1.885$W corresponding to $F_{3}$ is equal to $\hbar \omega_{d}|\bar{E}|^{2}$, where $|\bar{E}|^{2}$ is given in to Eq. (\ref{20}).  As an example, for $r = 3$ curve, the plot shows that $F_{3}=8.0842\times10^{-19}$N/$\sqrt{\mbox{Hz}}$ which matches exactly with the theoretical value calculated from Eq.(\ref{25}). It shows that the force sensitivity is improved by a factor of $20.05$ from the force sensitivity with no squeezing. For $r=0$ curve, the best sensitivity is same as $F_{2}$ which is the best possible sensitivity
when input states are not squeezed. If $r \to \infty$ then Eq. (\ref{25}) gives a force sensitivity of $ 3.86\times10^{-20}$N/$\sqrt{\mbox{Hz}}$ which is limited by the optomechanical mirror noise.\\
\\
Case (\romannumeral 2) : In Fig. \ref{diff_t}, the squeezing parameter is fixed
at $r = 3$ and four curves simulated by varying the squeezing
angle $\theta$. The parameters are same as in Fig. \ref{dip} so that
$1>\zeta^{2}/\Delta^{2}\gg\gamma\omega_{f}/\omega_{m}^{2}$, Eq. (\ref{16}) is satisfied and $\mathcal{R}(\alpha)\simeq\Delta$. So that RPN is suppressed. Figure. \ref{diff_t} clearly indicates that best sensitivity is achieved when $\theta=2\phi$ which
is in accordance with Eq. (\ref{23}). Again the best sensitivity for
$\theta=2\phi$ curve is exactly equal to $F_{3}$. For all other values of
$\theta$ the best sensitivity is smaller than $F_{3}$. The optical power
corresponding to the best sensitivity $F_{3}$ for $\theta=2\phi$ curve is related to Eq. (\ref{20})
as $\hbar \omega_{d}|\bar{E}|^{2}$. Overall both the Fig. \ref{diff_r} and Fig. \ref{diff_t} agree with the results in Eq. (\ref{23}) and Eq. (\ref{20}) and show that the improved force sensitivity is $\sqrt{e^{-2r}\zeta/4\Delta} F_{sql}$.

\par In this paragraph, we briefly describe some essential techniques for overcoming SQL. Quantum non-demolition measurements \cite{RevModPhys.68.1} can defeat SQL, but they require the measured variable to commute with the Hamiltonian. Coherent quantum noise cancellation \cite{PhysRevX.2.031016,Li2018} is another technique that overcomes
SQL by adding an auxiliary system to the main system. The auxiliary system parameters are tuned such that the quantum
back-action in the main system is cancelled. Variational measurement is used in Ref.\cite{PhysRevX.7.021008} to overcome SQL by using frequency dependent homodyne. Depending on the frequency region of interest, the phase of the reference beam is adjusted to
induce correlations between the amplitude and phase quadratures. Then these correlations are exploited to overcome SQL. Quantum mechanics free subsystems \cite{doi:10.1126/science.abf5389} overcomes SQL by
measuring combined quadratures like a sum of two amplitude
quadratures and the difference of the two phase quadratures. The
commutation relation of such combined quadratures is zero, leading to sub-SQL measurements. Signal recycling \cite{PhysRevD.64.042006,PhysRevD.67.062002,item_150019} is another technique where some of the output
from the interferometer is recycled back into the interferometer. An additional mirror achieves the recycling of output. This recycling mirror leads to additional resonances in
the interferometer, which can be adjusted to overcome SQL.
\section{Simulation parameters}
For simulation, we use the following optomechanical parameters: $m=10^{-7}$Kg, $\omega_{m}=10^{5}$Hz, $\omega_{f}=100$Hz, $\omega_{d}=6\pi/5\times10^{15}$Hz, $\Delta=100\zeta$, $\zeta=10^{6}$Hz,
$\gamma=1$Hz, $g=10^{18}$Hz/m. 
\section{Conclusion}
We discussed a new method to improve the sensitivity of squeezed light optomechanical interferometer for classical force detection. As a first step, we eliminated the RPN in an optomechanical cavity by using the quantum optical restoring force \cite{davuluri2022light}. In the next step, squeezed vacuum is sent through the empty input port of the interferometer. Combining the vacuum squeezing technique with RPN suppression through quantum optical restoring force improved the sensitivity of the squeezed light optomechanical interferometer. Generally, the vacuum squeezed light leads to classical force detection with sensitivity $e^{-r}F_{sql}$. In our method, the sensitivity is improved to $\sqrt{e^{-2r}\zeta/4\Delta}F_{sql}$, $1>\zeta/\Delta>0$, which is better than the sensitivity achieved by vacuum squeezing alone. We further studied dependence of force sensitivity on various input parameters. The method described in this article is applicable only to
frequencies much smaller than the resonance frequency of the
optomechanical mirror.
\section{Acknowledgements}
This work is supported by the Science and Engineering Research Board of India under the Grant no:SRG/2020/001167.
It is also supported by the National Natural Science Foundation of China (Grants No. 12074030 and No.
U1930402).
\bibliography{ref}

\begin{thebibliography}{78}%
\makeatletter
\providecommand \@ifxundefined [1]{%
 \@ifx{#1\undefined}
}%
\providecommand \@ifnum [1]{%
 \ifnum #1\expandafter \@firstoftwo
 \else \expandafter \@secondoftwo
 \fi
}%
\providecommand \@ifx [1]{%
 \ifx #1\expandafter \@firstoftwo
 \else \expandafter \@secondoftwo
 \fi
}%
\providecommand \natexlab [1]{#1}%
\providecommand \enquote  [1]{``#1''}%
\providecommand \bibnamefont  [1]{#1}%
\providecommand \bibfnamefont [1]{#1}%
\providecommand \citenamefont [1]{#1}%
\providecommand \href@noop [0]{\@secondoftwo}%
\providecommand \href [0]{\begingroup \@sanitize@url \@href}%
\providecommand \@href[1]{\@@startlink{#1}\@@href}%
\providecommand \@@href[1]{\endgroup#1\@@endlink}%
\providecommand \@sanitize@url [0]{\catcode `\\12\catcode `\$12\catcode
  `\&12\catcode `\#12\catcode `\^12\catcode `\_12\catcode `\%12\relax}%
\providecommand \@@startlink[1]{}%
\providecommand \@@endlink[0]{}%
\providecommand \url  [0]{\begingroup\@sanitize@url \@url }%
\providecommand \@url [1]{\endgroup\@href {#1}{\urlprefix }}%
\providecommand \urlprefix  [0]{URL }%
\providecommand \Eprint [0]{\href }%
\providecommand \doibase [0]{https://doi.org/}%
\providecommand \selectlanguage [0]{\@gobble}%
\providecommand \bibinfo  [0]{\@secondoftwo}%
\providecommand \bibfield  [0]{\@secondoftwo}%
\providecommand \translation [1]{[#1]}%
\providecommand \BibitemOpen [0]{}%
\providecommand \bibitemStop [0]{}%
\providecommand \bibitemNoStop [0]{.\EOS\space}%
\providecommand \EOS [0]{\spacefactor3000\relax}%
\providecommand \BibitemShut  [1]{\csname bibitem#1\endcsname}%
\let\auto@bib@innerbib\@empty
\bibitem [{\citenamefont {Giazotto}(1989)}]{GIAZOTTO1989365}%
  \BibitemOpen
  \bibfield  {author} {\bibinfo {author} {\bibfnamefont {A.}~\bibnamefont
  {Giazotto}},\ }\href
  {https://doi.org/https://doi.org/10.1016/0370-1573(89)90087-2} {\bibfield
  {journal} {\bibinfo  {journal} {Physics Reports}\ }\textbf {\bibinfo {volume}
  {182}},\ \bibinfo {pages} {365} (\bibinfo {year} {1989})}\BibitemShut
  {NoStop}%
\bibitem [{\citenamefont {Yu}\ \emph {et~al.}(2020)\citenamefont {Yu},
  \citenamefont {McCuller}, \citenamefont {Tse}, \citenamefont {Kijbunchoo},
  \citenamefont {Barsotti},\ and\ \citenamefont {\etal}}]{Yu2020}%
  \BibitemOpen
  \bibfield  {author} {\bibinfo {author} {\bibfnamefont {H.}~\bibnamefont
  {Yu}}, \bibinfo {author} {\bibfnamefont {L.}~\bibnamefont {McCuller}},
  \bibinfo {author} {\bibfnamefont {M.}~\bibnamefont {Tse}}, \bibinfo {author}
  {\bibfnamefont {N.}~\bibnamefont {Kijbunchoo}}, \bibinfo {author}
  {\bibfnamefont {L.}~\bibnamefont {Barsotti}},\ and\ \bibinfo {author}
  {\bibnamefont {\etal}},\ }\href {https://doi.org/10.1038/s41586-020-2420-8}
  {\bibfield  {journal} {\bibinfo  {journal} {Nature}\ }\textbf {\bibinfo
  {volume} {583}},\ \bibinfo {pages} {43} (\bibinfo {year} {2020})}\BibitemShut
  {NoStop}%
\bibitem [{\citenamefont {Meystre}(2012)}]{optomechanics1}%
  \BibitemOpen
  \bibfield  {author} {\bibinfo {author} {\bibfnamefont {P.}~\bibnamefont
  {Meystre}},\ }\href {https://doi.org/10.1002/andp.201200226} {\bibfield
  {journal} {\bibinfo  {journal} {Annalen der Physik}\ }\textbf {\bibinfo
  {volume} {525}},\ \bibinfo {pages} {215–233} (\bibinfo {year}
  {2012})}\BibitemShut {NoStop}%
\bibitem [{\citenamefont {Aspelmeyer}\ \emph {et~al.}(2014)\citenamefont
  {Aspelmeyer}, \citenamefont {Kippenberg},\ and\ \citenamefont
  {Marquardt}}]{RevModPhys.86.1391}%
  \BibitemOpen
  \bibfield  {author} {\bibinfo {author} {\bibfnamefont {M.}~\bibnamefont
  {Aspelmeyer}}, \bibinfo {author} {\bibfnamefont {T.~J.}\ \bibnamefont
  {Kippenberg}},\ and\ \bibinfo {author} {\bibfnamefont {F.}~\bibnamefont
  {Marquardt}},\ }\href {https://doi.org/10.1103/RevModPhys.86.1391} {\bibfield
   {journal} {\bibinfo  {journal} {Rev. Mod. Phys.}\ }\textbf {\bibinfo
  {volume} {86}},\ \bibinfo {pages} {1391} (\bibinfo {year}
  {2014})}\BibitemShut {NoStop}%
\bibitem [{\citenamefont {Kippenberg}\ and\ \citenamefont
  {Vahala}(2007)}]{Kippenberg:07}%
  \BibitemOpen
  \bibfield  {author} {\bibinfo {author} {\bibfnamefont {T.}~\bibnamefont
  {Kippenberg}}\ and\ \bibinfo {author} {\bibfnamefont {K.}~\bibnamefont
  {Vahala}},\ }\href {https://doi.org/10.1364/OE.15.017172} {\bibfield
  {journal} {\bibinfo  {journal} {Opt. Express}\ }\textbf {\bibinfo {volume}
  {15}},\ \bibinfo {pages} {17172} (\bibinfo {year} {2007})}\BibitemShut
  {NoStop}%
\bibitem [{\citenamefont {Barzanjeh}\ \emph {et~al.}(2022)\citenamefont
  {Barzanjeh}, \citenamefont {Xuereb}, \citenamefont {Gr{\"o}blacher},
  \citenamefont {Paternostro}, \citenamefont {Regal},\ and\ \citenamefont
  {Weig}}]{Barzanjeh2022}%
  \BibitemOpen
  \bibfield  {author} {\bibinfo {author} {\bibfnamefont {S.}~\bibnamefont
  {Barzanjeh}}, \bibinfo {author} {\bibfnamefont {A.}~\bibnamefont {Xuereb}},
  \bibinfo {author} {\bibfnamefont {S.}~\bibnamefont {Gr{\"o}blacher}},
  \bibinfo {author} {\bibfnamefont {M.}~\bibnamefont {Paternostro}}, \bibinfo
  {author} {\bibfnamefont {C.~A.}\ \bibnamefont {Regal}},\ and\ \bibinfo
  {author} {\bibfnamefont {E.~M.}\ \bibnamefont {Weig}},\ }\href
  {https://doi.org/10.1038/s41567-021-01402-0} {\bibfield  {journal} {\bibinfo
  {journal} {Nature Physics}\ }\textbf {\bibinfo {volume} {18}},\ \bibinfo
  {pages} {15} (\bibinfo {year} {2022})}\BibitemShut {NoStop}%
\bibitem [{\citenamefont {Arcizet}\ \emph {et~al.}(2006)\citenamefont
  {Arcizet}, \citenamefont {Cohadon}, \citenamefont {Briant}, \citenamefont
  {Pinard}, \citenamefont {Heidmann}, \citenamefont {Mackowski}, \citenamefont
  {Michel}, \citenamefont {Pinard}, \citenamefont
  {Fran\ifmmode~\mbox{\c{c}}\else \c{c}\fi{}ais},\ and\ \citenamefont
  {Rousseau}}]{PhysRevLett.97.133601}%
  \BibitemOpen
  \bibfield  {author} {\bibinfo {author} {\bibfnamefont {O.}~\bibnamefont
  {Arcizet}}, \bibinfo {author} {\bibfnamefont {P.-F.}\ \bibnamefont
  {Cohadon}}, \bibinfo {author} {\bibfnamefont {T.}~\bibnamefont {Briant}},
  \bibinfo {author} {\bibfnamefont {M.}~\bibnamefont {Pinard}}, \bibinfo
  {author} {\bibfnamefont {A.}~\bibnamefont {Heidmann}}, \bibinfo {author}
  {\bibfnamefont {J.-M.}\ \bibnamefont {Mackowski}}, \bibinfo {author}
  {\bibfnamefont {C.}~\bibnamefont {Michel}}, \bibinfo {author} {\bibfnamefont
  {L.}~\bibnamefont {Pinard}}, \bibinfo {author} {\bibfnamefont
  {O.}~\bibnamefont {Fran\ifmmode~\mbox{\c{c}}\else \c{c}\fi{}ais}},\ and\
  \bibinfo {author} {\bibfnamefont {L.}~\bibnamefont {Rousseau}},\ }\href
  {https://doi.org/10.1103/PhysRevLett.97.133601} {\bibfield  {journal}
  {\bibinfo  {journal} {Phys. Rev. Lett.}\ }\textbf {\bibinfo {volume} {97}},\
  \bibinfo {pages} {133601} (\bibinfo {year} {2006})}\BibitemShut {NoStop}%
\bibitem [{\citenamefont {LaHaye}\ \emph {et~al.}(2004)\citenamefont {LaHaye},
  \citenamefont {Buu}, \citenamefont {Camarota},\ and\ \citenamefont
  {Schwab}}]{miniaturizationOM3}%
  \BibitemOpen
  \bibfield  {author} {\bibinfo {author} {\bibfnamefont {M.~D.}\ \bibnamefont
  {LaHaye}}, \bibinfo {author} {\bibfnamefont {O.}~\bibnamefont {Buu}},
  \bibinfo {author} {\bibfnamefont {B.}~\bibnamefont {Camarota}},\ and\
  \bibinfo {author} {\bibfnamefont {K.~C.}\ \bibnamefont {Schwab}},\ }\href
  {https://doi.org/10.1126/science.1094419} {\bibfield  {journal} {\bibinfo
  {journal} {Science}\ }\textbf {\bibinfo {volume} {304}},\ \bibinfo {pages}
  {74} (\bibinfo {year} {2004})}\BibitemShut {NoStop}%
\bibitem [{\citenamefont {Pirkkalainen}\ \emph {et~al.}(2015)\citenamefont
  {Pirkkalainen}, \citenamefont {Cho}, \citenamefont {Massel}, \citenamefont
  {Tuorila}, \citenamefont {Heikkil{\"a}}, \citenamefont {Hakonen},\ and\
  \citenamefont {Sillanp{\"a}{\"a}}}]{cavityoptomechanics4nature}%
  \BibitemOpen
  \bibfield  {author} {\bibinfo {author} {\bibfnamefont {J.-M.}\ \bibnamefont
  {Pirkkalainen}}, \bibinfo {author} {\bibfnamefont {S.}~\bibnamefont {Cho}},
  \bibinfo {author} {\bibfnamefont {F.}~\bibnamefont {Massel}}, \bibinfo
  {author} {\bibfnamefont {J.}~\bibnamefont {Tuorila}}, \bibinfo {author}
  {\bibfnamefont {T.}~\bibnamefont {Heikkil{\"a}}}, \bibinfo {author}
  {\bibfnamefont {P.}~\bibnamefont {Hakonen}},\ and\ \bibinfo {author}
  {\bibfnamefont {M.}~\bibnamefont {Sillanp{\"a}{\"a}}},\ }\href
  {https://doi.org/10.1038/ncomms7981} {\bibfield  {journal} {\bibinfo
  {journal} {Nature communications}\ }\textbf {\bibinfo {volume} {6}},\
  \bibinfo {pages} {1} (\bibinfo {year} {2015})}\BibitemShut {NoStop}%
\bibitem [{\citenamefont {Li}\ \emph {et~al.}(2021)\citenamefont {Li},
  \citenamefont {Ou}, \citenamefont {Lei},\ and\ \citenamefont
  {Liu}}]{sensors}%
  \BibitemOpen
  \bibfield  {author} {\bibinfo {author} {\bibfnamefont {B.-B.}\ \bibnamefont
  {Li}}, \bibinfo {author} {\bibfnamefont {L.}~\bibnamefont {Ou}}, \bibinfo
  {author} {\bibfnamefont {Y.}~\bibnamefont {Lei}},\ and\ \bibinfo {author}
  {\bibfnamefont {Y.-C.}\ \bibnamefont {Liu}},\ }\href
  {https://doi.org/doi:10.1515/nanoph-2021-0256} {\bibfield  {journal}
  {\bibinfo  {journal} {Nanophotonics}\ }\textbf {\bibinfo {volume} {10}},\
  \bibinfo {pages} {2799} (\bibinfo {year} {2021})}\BibitemShut {NoStop}%
\bibitem [{\citenamefont {Moser}\ \emph {et~al.}(2013)\citenamefont {Moser},
  \citenamefont {G{\"u}ttinger}, \citenamefont {Eichler}, \citenamefont
  {Esplandiu}, \citenamefont {Liu}, \citenamefont {Dykman},\ and\ \citenamefont
  {Bachtold}}]{forcesensor}%
  \BibitemOpen
  \bibfield  {author} {\bibinfo {author} {\bibfnamefont {J.}~\bibnamefont
  {Moser}}, \bibinfo {author} {\bibfnamefont {J.}~\bibnamefont
  {G{\"u}ttinger}}, \bibinfo {author} {\bibfnamefont {A.}~\bibnamefont
  {Eichler}}, \bibinfo {author} {\bibfnamefont {M.~J.}\ \bibnamefont
  {Esplandiu}}, \bibinfo {author} {\bibfnamefont {D.~E.}\ \bibnamefont {Liu}},
  \bibinfo {author} {\bibfnamefont {M.~I.}\ \bibnamefont {Dykman}},\ and\
  \bibinfo {author} {\bibfnamefont {A.}~\bibnamefont {Bachtold}},\ }\href
  {https://doi.org/10.1038/nnano.2013.97} {\bibfield  {journal} {\bibinfo
  {journal} {Nature Nanotechnology}\ }\textbf {\bibinfo {volume} {8}},\
  \bibinfo {pages} {493} (\bibinfo {year} {2013})}\BibitemShut {NoStop}%
\bibitem [{\citenamefont {Bocko}\ and\ \citenamefont
  {Onofrio}(1996)}]{RevModPhys.68.755}%
  \BibitemOpen
  \bibfield  {author} {\bibinfo {author} {\bibfnamefont {M.~F.}\ \bibnamefont
  {Bocko}}\ and\ \bibinfo {author} {\bibfnamefont {R.}~\bibnamefont
  {Onofrio}},\ }\href {https://doi.org/10.1103/RevModPhys.68.755} {\bibfield
  {journal} {\bibinfo  {journal} {Rev. Mod. Phys.}\ }\textbf {\bibinfo {volume}
  {68}},\ \bibinfo {pages} {755} (\bibinfo {year} {1996})}\BibitemShut
  {NoStop}%
\bibitem [{\citenamefont {Caves}\ \emph {et~al.}(1980)\citenamefont {Caves},
  \citenamefont {Thorne}, \citenamefont {Drever}, \citenamefont {Sandberg},\
  and\ \citenamefont {Zimmermann}}]{RevModPhys.52.341}%
  \BibitemOpen
  \bibfield  {author} {\bibinfo {author} {\bibfnamefont {C.~M.}\ \bibnamefont
  {Caves}}, \bibinfo {author} {\bibfnamefont {K.~S.}\ \bibnamefont {Thorne}},
  \bibinfo {author} {\bibfnamefont {R.~W.~P.}\ \bibnamefont {Drever}}, \bibinfo
  {author} {\bibfnamefont {V.~D.}\ \bibnamefont {Sandberg}},\ and\ \bibinfo
  {author} {\bibfnamefont {M.}~\bibnamefont {Zimmermann}},\ }\href
  {https://doi.org/10.1103/RevModPhys.52.341} {\bibfield  {journal} {\bibinfo
  {journal} {Rev. Mod. Phys.}\ }\textbf {\bibinfo {volume} {52}},\ \bibinfo
  {pages} {341} (\bibinfo {year} {1980})}\BibitemShut {NoStop}%
\bibitem [{\citenamefont {Komori}\ \emph {et~al.}(2021)\citenamefont {Komori},
  \citenamefont {Kawasaki}, \citenamefont {Otabe}, \citenamefont {Enomoto},
  \citenamefont {Michimura},\ and\ \citenamefont
  {Ando}}]{PhysRevA.104.L031501}%
  \BibitemOpen
  \bibfield  {author} {\bibinfo {author} {\bibfnamefont {K.}~\bibnamefont
  {Komori}}, \bibinfo {author} {\bibfnamefont {T.}~\bibnamefont {Kawasaki}},
  \bibinfo {author} {\bibfnamefont {S.}~\bibnamefont {Otabe}}, \bibinfo
  {author} {\bibfnamefont {Y.}~\bibnamefont {Enomoto}}, \bibinfo {author}
  {\bibfnamefont {Y.}~\bibnamefont {Michimura}},\ and\ \bibinfo {author}
  {\bibfnamefont {M.}~\bibnamefont {Ando}},\ }\href
  {https://doi.org/10.1103/PhysRevA.104.L031501} {\bibfield  {journal}
  {\bibinfo  {journal} {Phys. Rev. A}\ }\textbf {\bibinfo {volume} {104}},\
  \bibinfo {pages} {L031501} (\bibinfo {year} {2021})}\BibitemShut {NoStop}%
\bibitem [{\citenamefont {Cripe}\ \emph {et~al.}(2020)\citenamefont {Cripe},
  \citenamefont {Cullen}, \citenamefont {Chen}, \citenamefont {Heu},
  \citenamefont {Follman}, \citenamefont {Cole},\ and\ \citenamefont
  {Corbitt}}]{PhysRevX.10.031065}%
  \BibitemOpen
  \bibfield  {author} {\bibinfo {author} {\bibfnamefont {J.}~\bibnamefont
  {Cripe}}, \bibinfo {author} {\bibfnamefont {T.}~\bibnamefont {Cullen}},
  \bibinfo {author} {\bibfnamefont {Y.}~\bibnamefont {Chen}}, \bibinfo {author}
  {\bibfnamefont {P.}~\bibnamefont {Heu}}, \bibinfo {author} {\bibfnamefont
  {D.}~\bibnamefont {Follman}}, \bibinfo {author} {\bibfnamefont {G.~D.}\
  \bibnamefont {Cole}},\ and\ \bibinfo {author} {\bibfnamefont
  {T.}~\bibnamefont {Corbitt}},\ }\href
  {https://doi.org/10.1103/PhysRevX.10.031065} {\bibfield  {journal} {\bibinfo
  {journal} {Phys. Rev. X}\ }\textbf {\bibinfo {volume} {10}},\ \bibinfo
  {pages} {031065} (\bibinfo {year} {2020})}\BibitemShut {NoStop}%
\bibitem [{\citenamefont {Edelstein}\ \emph {et~al.}(1978)\citenamefont
  {Edelstein}, \citenamefont {Hough}, \citenamefont {Pugh},\ and\ \citenamefont
  {Martin}}]{LIGO2}%
  \BibitemOpen
  \bibfield  {author} {\bibinfo {author} {\bibfnamefont {W.~A.}\ \bibnamefont
  {Edelstein}}, \bibinfo {author} {\bibfnamefont {J.}~\bibnamefont {Hough}},
  \bibinfo {author} {\bibfnamefont {J.~R.}\ \bibnamefont {Pugh}},\ and\
  \bibinfo {author} {\bibfnamefont {W.}~\bibnamefont {Martin}},\ }\href
  {http://stacks.iop.org/0022-3735/11/i=7/a=030} {\bibfield  {journal}
  {\bibinfo  {journal} {Journal of Physics E: Scientific Instruments}\ }\textbf
  {\bibinfo {volume} {11}},\ \bibinfo {pages} {710} (\bibinfo {year}
  {1978})}\BibitemShut {NoStop}%
\bibitem [{\citenamefont {Caves}(1980)}]{PhysRevLett.45.75}%
  \BibitemOpen
  \bibfield  {author} {\bibinfo {author} {\bibfnamefont {C.~M.}\ \bibnamefont
  {Caves}},\ }\href {https://doi.org/10.1103/PhysRevLett.45.75} {\bibfield
  {journal} {\bibinfo  {journal} {Phys. Rev. Lett.}\ }\textbf {\bibinfo
  {volume} {45}},\ \bibinfo {pages} {75} (\bibinfo {year} {1980})}\BibitemShut
  {NoStop}%
\bibitem [{\citenamefont {Murch}\ \emph {et~al.}(2008)\citenamefont {Murch},
  \citenamefont {Moore}, \citenamefont {Gupta},\ and\ \citenamefont
  {Stamper-Kurn}}]{Murch2008}%
  \BibitemOpen
  \bibfield  {author} {\bibinfo {author} {\bibfnamefont {K.~W.}\ \bibnamefont
  {Murch}}, \bibinfo {author} {\bibfnamefont {K.~L.}\ \bibnamefont {Moore}},
  \bibinfo {author} {\bibfnamefont {S.}~\bibnamefont {Gupta}},\ and\ \bibinfo
  {author} {\bibfnamefont {D.~M.}\ \bibnamefont {Stamper-Kurn}},\ }\href
  {https://doi.org/10.1038/nphys965} {\bibfield  {journal} {\bibinfo  {journal}
  {Nature Physics}\ }\textbf {\bibinfo {volume} {4}},\ \bibinfo {pages} {561}
  (\bibinfo {year} {2008})}\BibitemShut {NoStop}%
\bibitem [{\citenamefont {Cripe}\ \emph {et~al.}(2019)\citenamefont {Cripe},
  \citenamefont {Aggarwal}, \citenamefont {Lanza}, \citenamefont {Libson},
  \citenamefont {Singh}, \citenamefont {Heu}, \citenamefont {Follman},
  \citenamefont {Cole}, \citenamefont {Mavalvala},\ and\ \citenamefont
  {Corbitt}}]{Cripe2019}%
  \BibitemOpen
  \bibfield  {author} {\bibinfo {author} {\bibfnamefont {J.}~\bibnamefont
  {Cripe}}, \bibinfo {author} {\bibfnamefont {N.}~\bibnamefont {Aggarwal}},
  \bibinfo {author} {\bibfnamefont {R.}~\bibnamefont {Lanza}}, \bibinfo
  {author} {\bibfnamefont {A.}~\bibnamefont {Libson}}, \bibinfo {author}
  {\bibfnamefont {R.}~\bibnamefont {Singh}}, \bibinfo {author} {\bibfnamefont
  {P.}~\bibnamefont {Heu}}, \bibinfo {author} {\bibfnamefont {D.}~\bibnamefont
  {Follman}}, \bibinfo {author} {\bibfnamefont {G.~D.}\ \bibnamefont {Cole}},
  \bibinfo {author} {\bibfnamefont {N.}~\bibnamefont {Mavalvala}},\ and\
  \bibinfo {author} {\bibfnamefont {T.}~\bibnamefont {Corbitt}},\ }\href
  {https://doi.org/10.1038/s41586-019-1051-4} {\bibfield  {journal} {\bibinfo
  {journal} {Nature}\ }\textbf {\bibinfo {volume} {568}},\ \bibinfo {pages}
  {364} (\bibinfo {year} {2019})}\BibitemShut {NoStop}%
\bibitem [{\citenamefont {Davuluri}(2016)}]{PhysRevA.94.013808}%
  \BibitemOpen
  \bibfield  {author} {\bibinfo {author} {\bibfnamefont {S.}~\bibnamefont
  {Davuluri}},\ }\href {https://doi.org/10.1103/PhysRevA.94.013808} {\bibfield
  {journal} {\bibinfo  {journal} {Phys. Rev. A}\ }\textbf {\bibinfo {volume}
  {94}},\ \bibinfo {pages} {013808} (\bibinfo {year} {2016})}\BibitemShut
  {NoStop}%
\bibitem [{\citenamefont {Bondurant}(1986)}]{PhysRevA.34.3927}%
  \BibitemOpen
  \bibfield  {author} {\bibinfo {author} {\bibfnamefont {R.~S.}\ \bibnamefont
  {Bondurant}},\ }\href {https://doi.org/10.1103/PhysRevA.34.3927} {\bibfield
  {journal} {\bibinfo  {journal} {Phys. Rev. A}\ }\textbf {\bibinfo {volume}
  {34}},\ \bibinfo {pages} {3927} (\bibinfo {year} {1986})}\BibitemShut
  {NoStop}%
\bibitem [{\citenamefont {Braginsky}\ \emph {et~al.}(1980)\citenamefont
  {Braginsky}, \citenamefont {Vorontsov},\ and\ \citenamefont
  {Thorne}}]{braginsky1980quantum}%
  \BibitemOpen
  \bibfield  {author} {\bibinfo {author} {\bibfnamefont {V.~B.}\ \bibnamefont
  {Braginsky}}, \bibinfo {author} {\bibfnamefont {Y.~I.}\ \bibnamefont
  {Vorontsov}},\ and\ \bibinfo {author} {\bibfnamefont {K.~S.}\ \bibnamefont
  {Thorne}},\ }\href {https://doi.org/10.1126/science.209.4456.547} {\bibfield
  {journal} {\bibinfo  {journal} {Science}\ }\textbf {\bibinfo {volume}
  {209}},\ \bibinfo {pages} {547} (\bibinfo {year} {1980})}\BibitemShut
  {NoStop}%
\bibitem [{\citenamefont {Hertzberg}\ \emph {et~al.}(2010)\citenamefont
  {Hertzberg}, \citenamefont {Rocheleau}, \citenamefont {Ndukum}, \citenamefont
  {Savva}, \citenamefont {Clerk},\ and\ \citenamefont {Schwab}}]{QND2}%
  \BibitemOpen
  \bibfield  {author} {\bibinfo {author} {\bibfnamefont {J.~B.}\ \bibnamefont
  {Hertzberg}}, \bibinfo {author} {\bibfnamefont {T.}~\bibnamefont
  {Rocheleau}}, \bibinfo {author} {\bibfnamefont {T.}~\bibnamefont {Ndukum}},
  \bibinfo {author} {\bibfnamefont {M.}~\bibnamefont {Savva}}, \bibinfo
  {author} {\bibfnamefont {A.~A.}\ \bibnamefont {Clerk}},\ and\ \bibinfo
  {author} {\bibfnamefont {K.~C.}\ \bibnamefont {Schwab}},\ }\href
  {https://doi.org/10.1038/nphys1479} {\bibfield  {journal} {\bibinfo
  {journal} {Nature Physics}\ }\textbf {\bibinfo {volume} {6}},\ \bibinfo
  {pages} {213} (\bibinfo {year} {2010})}\BibitemShut {NoStop}%
\bibitem [{\citenamefont {Suh}\ \emph {et~al.}(2014)\citenamefont {Suh},
  \citenamefont {Weinstein}, \citenamefont {Lei}, \citenamefont {Wollman},
  \citenamefont {Steinke}, \citenamefont {Meystre}, \citenamefont {Clerk},\
  and\ \citenamefont {Schwab}}]{QND3}%
  \BibitemOpen
  \bibfield  {author} {\bibinfo {author} {\bibfnamefont {J.}~\bibnamefont
  {Suh}}, \bibinfo {author} {\bibfnamefont {A.~J.}\ \bibnamefont {Weinstein}},
  \bibinfo {author} {\bibfnamefont {C.~U.}\ \bibnamefont {Lei}}, \bibinfo
  {author} {\bibfnamefont {E.~E.}\ \bibnamefont {Wollman}}, \bibinfo {author}
  {\bibfnamefont {S.~K.}\ \bibnamefont {Steinke}}, \bibinfo {author}
  {\bibfnamefont {P.}~\bibnamefont {Meystre}}, \bibinfo {author} {\bibfnamefont
  {A.~A.}\ \bibnamefont {Clerk}},\ and\ \bibinfo {author} {\bibfnamefont
  {K.~C.}\ \bibnamefont {Schwab}},\ }\href
  {https://doi.org/10.1126/science.1253258} {\bibfield  {journal} {\bibinfo
  {journal} {Science}\ }\textbf {\bibinfo {volume} {344}},\ \bibinfo {pages}
  {1262} (\bibinfo {year} {2014})}\BibitemShut {NoStop}%
\bibitem [{\citenamefont {Thorne}\ \emph {et~al.}(1978)\citenamefont {Thorne},
  \citenamefont {Drever}, \citenamefont {Caves}, \citenamefont {Zimmermann},\
  and\ \citenamefont {Sandberg}}]{PhysRevLett.40.667}%
  \BibitemOpen
  \bibfield  {author} {\bibinfo {author} {\bibfnamefont {K.~S.}\ \bibnamefont
  {Thorne}}, \bibinfo {author} {\bibfnamefont {R.~W.~P.}\ \bibnamefont
  {Drever}}, \bibinfo {author} {\bibfnamefont {C.~M.}\ \bibnamefont {Caves}},
  \bibinfo {author} {\bibfnamefont {M.}~\bibnamefont {Zimmermann}},\ and\
  \bibinfo {author} {\bibfnamefont {V.~D.}\ \bibnamefont {Sandberg}},\ }\href
  {https://doi.org/10.1103/PhysRevLett.40.667} {\bibfield  {journal} {\bibinfo
  {journal} {Phys. Rev. Lett.}\ }\textbf {\bibinfo {volume} {40}},\ \bibinfo
  {pages} {667} (\bibinfo {year} {1978})}\BibitemShut {NoStop}%
\bibitem [{\citenamefont {Clerk}\ \emph {et~al.}(2008)\citenamefont {Clerk},
  \citenamefont {Marquardt},\ and\ \citenamefont {Jacobs}}]{Clerk_2008}%
  \BibitemOpen
  \bibfield  {author} {\bibinfo {author} {\bibfnamefont {A.~A.}\ \bibnamefont
  {Clerk}}, \bibinfo {author} {\bibfnamefont {F.}~\bibnamefont {Marquardt}},\
  and\ \bibinfo {author} {\bibfnamefont {K.}~\bibnamefont {Jacobs}},\ }\href
  {https://doi.org/10.1088/1367-2630/10/9/095010} {\bibfield  {journal}
  {\bibinfo  {journal} {New Journal of Physics}\ }\textbf {\bibinfo {volume}
  {10}},\ \bibinfo {pages} {095010} (\bibinfo {year} {2008})}\BibitemShut
  {NoStop}%
\bibitem [{\citenamefont {Vyatchanin}\ and\ \citenamefont
  {Zubova}(1995)}]{variationalmeasurement}%
  \BibitemOpen
  \bibfield  {author} {\bibinfo {author} {\bibfnamefont {S.}~\bibnamefont
  {Vyatchanin}}\ and\ \bibinfo {author} {\bibfnamefont {E.}~\bibnamefont
  {Zubova}},\ }\href
  {https://doi.org/https://doi.org/10.1016/0375-9601(95)00280-G} {\bibfield
  {journal} {\bibinfo  {journal} {Physics Letters A}\ }\textbf {\bibinfo
  {volume} {201}},\ \bibinfo {pages} {269} (\bibinfo {year}
  {1995})}\BibitemShut {NoStop}%
\bibitem [{\citenamefont {Tsang}\ and\ \citenamefont
  {Caves}(2010)}]{PhysRevLett.105.123601}%
  \BibitemOpen
  \bibfield  {author} {\bibinfo {author} {\bibfnamefont {M.}~\bibnamefont
  {Tsang}}\ and\ \bibinfo {author} {\bibfnamefont {C.~M.}\ \bibnamefont
  {Caves}},\ }\href {https://doi.org/10.1103/PhysRevLett.105.123601} {\bibfield
   {journal} {\bibinfo  {journal} {Phys. Rev. Lett.}\ }\textbf {\bibinfo
  {volume} {105}},\ \bibinfo {pages} {123601} (\bibinfo {year}
  {2010})}\BibitemShut {NoStop}%
\bibitem [{\citenamefont {M{\o}ller}\ \emph {et~al.}(2017)\citenamefont
  {M{\o}ller}, \citenamefont {Thomas}, \citenamefont {Vasilakis}, \citenamefont
  {Zeuthen}, \citenamefont {Tsaturyan}, \citenamefont {Balabas}, \citenamefont
  {Jensen}, \citenamefont {Schliesser}, \citenamefont {Hammerer},\ and\
  \citenamefont {Polzik}}]{negativemassreferenceframe}%
  \BibitemOpen
  \bibfield  {author} {\bibinfo {author} {\bibfnamefont {C.~B.}\ \bibnamefont
  {M{\o}ller}}, \bibinfo {author} {\bibfnamefont {R.~A.}\ \bibnamefont
  {Thomas}}, \bibinfo {author} {\bibfnamefont {G.}~\bibnamefont {Vasilakis}},
  \bibinfo {author} {\bibfnamefont {E.}~\bibnamefont {Zeuthen}}, \bibinfo
  {author} {\bibfnamefont {Y.}~\bibnamefont {Tsaturyan}}, \bibinfo {author}
  {\bibfnamefont {M.}~\bibnamefont {Balabas}}, \bibinfo {author} {\bibfnamefont
  {K.}~\bibnamefont {Jensen}}, \bibinfo {author} {\bibfnamefont
  {A.}~\bibnamefont {Schliesser}}, \bibinfo {author} {\bibfnamefont
  {K.}~\bibnamefont {Hammerer}},\ and\ \bibinfo {author} {\bibfnamefont
  {E.~S.}\ \bibnamefont {Polzik}},\ }\href
  {https://doi.org/10.1038/nature22980} {\bibfield  {journal} {\bibinfo
  {journal} {Nature}\ }\textbf {\bibinfo {volume} {547}},\ \bibinfo {pages}
  {191} (\bibinfo {year} {2017})}\BibitemShut {NoStop}%
\bibitem [{\citenamefont {Giovannetti}\ \emph {et~al.}(2004)\citenamefont
  {Giovannetti}, \citenamefont {Lloyd},\ and\ \citenamefont
  {Maccone}}]{doi:10.1126/science.1104149}%
  \BibitemOpen
  \bibfield  {author} {\bibinfo {author} {\bibfnamefont {V.}~\bibnamefont
  {Giovannetti}}, \bibinfo {author} {\bibfnamefont {S.}~\bibnamefont {Lloyd}},\
  and\ \bibinfo {author} {\bibfnamefont {L.}~\bibnamefont {Maccone}},\ }\href
  {https://doi.org/10.1126/science.1104149} {\bibfield  {journal} {\bibinfo
  {journal} {Science}\ }\textbf {\bibinfo {volume} {306}},\ \bibinfo {pages}
  {1330} (\bibinfo {year} {2004})}\BibitemShut {NoStop}%
\bibitem [{\citenamefont {Xie}\ \emph {et~al.}(2021)\citenamefont {Xie},
  \citenamefont {Zhao}, \citenamefont {Kong}, \citenamefont {Ma}, \citenamefont
  {Wang}, \citenamefont {Ye}, \citenamefont {Yu}, \citenamefont {Yang},
  \citenamefont {Xu}, \citenamefont {Wang}, \citenamefont {Wang}, \citenamefont
  {Shi},\ and\ \citenamefont {Du}}]{doi:10.1126/sciadv.abg9204}%
  \BibitemOpen
  \bibfield  {author} {\bibinfo {author} {\bibfnamefont {T.}~\bibnamefont
  {Xie}}, \bibinfo {author} {\bibfnamefont {Z.}~\bibnamefont {Zhao}}, \bibinfo
  {author} {\bibfnamefont {X.}~\bibnamefont {Kong}}, \bibinfo {author}
  {\bibfnamefont {W.}~\bibnamefont {Ma}}, \bibinfo {author} {\bibfnamefont
  {M.}~\bibnamefont {Wang}}, \bibinfo {author} {\bibfnamefont {X.}~\bibnamefont
  {Ye}}, \bibinfo {author} {\bibfnamefont {P.}~\bibnamefont {Yu}}, \bibinfo
  {author} {\bibfnamefont {Z.}~\bibnamefont {Yang}}, \bibinfo {author}
  {\bibfnamefont {S.}~\bibnamefont {Xu}}, \bibinfo {author} {\bibfnamefont
  {P.}~\bibnamefont {Wang}}, \bibinfo {author} {\bibfnamefont {Y.}~\bibnamefont
  {Wang}}, \bibinfo {author} {\bibfnamefont {F.}~\bibnamefont {Shi}},\ and\
  \bibinfo {author} {\bibfnamefont {J.}~\bibnamefont {Du}},\ }\href
  {https://doi.org/10.1126/sciadv.abg9204} {\bibfield  {journal} {\bibinfo
  {journal} {Science Advances}\ }\textbf {\bibinfo {volume} {7}},\ \bibinfo
  {pages} {eabg9204} (\bibinfo {year} {2021})}\BibitemShut {NoStop}%
\bibitem [{\citenamefont {Nagata}\ \emph {et~al.}(2007)\citenamefont {Nagata},
  \citenamefont {Okamoto}, \citenamefont {O'Brien}, \citenamefont {Sasaki},\
  and\ \citenamefont {Takeuchi}}]{doi:10.1126/science.1138007}%
  \BibitemOpen
  \bibfield  {author} {\bibinfo {author} {\bibfnamefont {T.}~\bibnamefont
  {Nagata}}, \bibinfo {author} {\bibfnamefont {R.}~\bibnamefont {Okamoto}},
  \bibinfo {author} {\bibfnamefont {J.~L.}\ \bibnamefont {O'Brien}}, \bibinfo
  {author} {\bibfnamefont {K.}~\bibnamefont {Sasaki}},\ and\ \bibinfo {author}
  {\bibfnamefont {S.}~\bibnamefont {Takeuchi}},\ }\href
  {https://doi.org/10.1126/science.1138007} {\bibfield  {journal} {\bibinfo
  {journal} {Science}\ }\textbf {\bibinfo {volume} {316}},\ \bibinfo {pages}
  {726} (\bibinfo {year} {2007})}\BibitemShut {NoStop}%
\bibitem [{\citenamefont {Buchmann}\ \emph {et~al.}(2016)\citenamefont
  {Buchmann}, \citenamefont {Schreppler}, \citenamefont {Kohler}, \citenamefont
  {Spethmann},\ and\ \citenamefont {Stamper-Kurn}}]{PhysRevLett.117.030801}%
  \BibitemOpen
  \bibfield  {author} {\bibinfo {author} {\bibfnamefont {L.~F.}\ \bibnamefont
  {Buchmann}}, \bibinfo {author} {\bibfnamefont {S.}~\bibnamefont
  {Schreppler}}, \bibinfo {author} {\bibfnamefont {J.}~\bibnamefont {Kohler}},
  \bibinfo {author} {\bibfnamefont {N.}~\bibnamefont {Spethmann}},\ and\
  \bibinfo {author} {\bibfnamefont {D.~M.}\ \bibnamefont {Stamper-Kurn}},\
  }\href {https://doi.org/10.1103/PhysRevLett.117.030801} {\bibfield  {journal}
  {\bibinfo  {journal} {Phys. Rev. Lett.}\ }\textbf {\bibinfo {volume} {117}},\
  \bibinfo {pages} {030801} (\bibinfo {year} {2016})}\BibitemShut {NoStop}%
\bibitem [{\citenamefont {Ockeloen-Korppi}\ \emph {et~al.}(2018)\citenamefont
  {Ockeloen-Korppi}, \citenamefont {Damsk\"agg}, \citenamefont {Paraoanu},
  \citenamefont {Massel},\ and\ \citenamefont
  {Sillanp\"a\"a}}]{PhysRevLett.121.243601}%
  \BibitemOpen
  \bibfield  {author} {\bibinfo {author} {\bibfnamefont {C.~F.}\ \bibnamefont
  {Ockeloen-Korppi}}, \bibinfo {author} {\bibfnamefont {E.}~\bibnamefont
  {Damsk\"agg}}, \bibinfo {author} {\bibfnamefont {G.~S.}\ \bibnamefont
  {Paraoanu}}, \bibinfo {author} {\bibfnamefont {F.}~\bibnamefont {Massel}},\
  and\ \bibinfo {author} {\bibfnamefont {M.~A.}\ \bibnamefont
  {Sillanp\"a\"a}},\ }\href {https://doi.org/10.1103/PhysRevLett.121.243601}
  {\bibfield  {journal} {\bibinfo  {journal} {Phys. Rev. Lett.}\ }\textbf
  {\bibinfo {volume} {121}},\ \bibinfo {pages} {243601} (\bibinfo {year}
  {2018})}\BibitemShut {NoStop}%
\bibitem [{\citenamefont {Jaekel}\ and\ \citenamefont
  {Reynaud}(1990)}]{Jaekel_1990}%
  \BibitemOpen
  \bibfield  {author} {\bibinfo {author} {\bibfnamefont {M.~T.}\ \bibnamefont
  {Jaekel}}\ and\ \bibinfo {author} {\bibfnamefont {S.}~\bibnamefont
  {Reynaud}},\ }\href {https://doi.org/10.1209/0295-5075/13/4/003} {\bibfield
  {journal} {\bibinfo  {journal} {Europhysics Letters ({EPL})}\ }\textbf
  {\bibinfo {volume} {13}},\ \bibinfo {pages} {301} (\bibinfo {year}
  {1990})}\BibitemShut {NoStop}%
\bibitem [{\citenamefont {Ma}\ \emph {et~al.}(2017)\citenamefont {Ma},
  \citenamefont {Miao}, \citenamefont {Pang}, \citenamefont {Evans},
  \citenamefont {Zhao}, \citenamefont {Harms}, \citenamefont {Schnabel},\ and\
  \citenamefont {Chen}}]{Ma2017}%
  \BibitemOpen
  \bibfield  {author} {\bibinfo {author} {\bibfnamefont {Y.}~\bibnamefont
  {Ma}}, \bibinfo {author} {\bibfnamefont {H.}~\bibnamefont {Miao}}, \bibinfo
  {author} {\bibfnamefont {B.~H.}\ \bibnamefont {Pang}}, \bibinfo {author}
  {\bibfnamefont {M.}~\bibnamefont {Evans}}, \bibinfo {author} {\bibfnamefont
  {C.}~\bibnamefont {Zhao}}, \bibinfo {author} {\bibfnamefont {J.}~\bibnamefont
  {Harms}}, \bibinfo {author} {\bibfnamefont {R.}~\bibnamefont {Schnabel}},\
  and\ \bibinfo {author} {\bibfnamefont {Y.}~\bibnamefont {Chen}},\ }\href
  {https://doi.org/10.1038/nphys4118} {\bibfield  {journal} {\bibinfo
  {journal} {Nature Physics}\ }\textbf {\bibinfo {volume} {13}},\ \bibinfo
  {pages} {776} (\bibinfo {year} {2017})}\BibitemShut {NoStop}%
\bibitem [{\citenamefont {Schnabel}(2017)}]{squeezing2}%
  \BibitemOpen
  \bibfield  {author} {\bibinfo {author} {\bibfnamefont {R.}~\bibnamefont
  {Schnabel}},\ }\href
  {https://doi.org/https://doi.org/10.1016/j.physrep.2017.04.001} {\bibfield
  {journal} {\bibinfo  {journal} {Physics Reports}\ }\textbf {\bibinfo {volume}
  {684}},\ \bibinfo {pages} {1} (\bibinfo {year} {2017})}\BibitemShut {NoStop}%
\bibitem [{\citenamefont {Lawrie}\ \emph {et~al.}(2019)\citenamefont {Lawrie},
  \citenamefont {Lett}, \citenamefont {Marino},\ and\ \citenamefont
  {Pooser}}]{Lawrie2019}%
  \BibitemOpen
  \bibfield  {author} {\bibinfo {author} {\bibfnamefont {B.~J.}\ \bibnamefont
  {Lawrie}}, \bibinfo {author} {\bibfnamefont {P.~D.}\ \bibnamefont {Lett}},
  \bibinfo {author} {\bibfnamefont {A.~M.}\ \bibnamefont {Marino}},\ and\
  \bibinfo {author} {\bibfnamefont {R.~C.}\ \bibnamefont {Pooser}},\ }\href
  {https://doi.org/10.1021/acsphotonics.9b00250} {\bibfield  {journal}
  {\bibinfo  {journal} {ACS Photonics}\ }\textbf {\bibinfo {volume} {6}},\
  \bibinfo {pages} {1307} (\bibinfo {year} {2019})}\BibitemShut {NoStop}%
\bibitem [{\citenamefont {Lee}\ \emph {et~al.}(2020)\citenamefont {Lee},
  \citenamefont {Lee},\ and\ \citenamefont {Seok}}]{Lee2020}%
  \BibitemOpen
  \bibfield  {author} {\bibinfo {author} {\bibfnamefont {C.-W.}\ \bibnamefont
  {Lee}}, \bibinfo {author} {\bibfnamefont {J.~H.}\ \bibnamefont {Lee}},\ and\
  \bibinfo {author} {\bibfnamefont {H.}~\bibnamefont {Seok}},\ }\href
  {https://doi.org/10.1038/s41598-020-74629-1} {\bibfield  {journal} {\bibinfo
  {journal} {Scientific Reports}\ }\textbf {\bibinfo {volume} {10}},\ \bibinfo
  {pages} {17496} (\bibinfo {year} {2020})}\BibitemShut {NoStop}%
\bibitem [{\citenamefont {Aggarwal}\ \emph {et~al.}(2020)\citenamefont
  {Aggarwal}, \citenamefont {Cullen}, \citenamefont {Cripe}, \citenamefont
  {Cole}, \citenamefont {Lanza}, \citenamefont {Libson}, \citenamefont
  {Follman}, \citenamefont {Heu}, \citenamefont {Corbitt},\ and\ \citenamefont
  {Mavalvala}}]{Aggarwal2020}%
  \BibitemOpen
  \bibfield  {author} {\bibinfo {author} {\bibfnamefont {N.}~\bibnamefont
  {Aggarwal}}, \bibinfo {author} {\bibfnamefont {T.~J.}\ \bibnamefont
  {Cullen}}, \bibinfo {author} {\bibfnamefont {J.}~\bibnamefont {Cripe}},
  \bibinfo {author} {\bibfnamefont {G.~D.}\ \bibnamefont {Cole}}, \bibinfo
  {author} {\bibfnamefont {R.}~\bibnamefont {Lanza}}, \bibinfo {author}
  {\bibfnamefont {A.}~\bibnamefont {Libson}}, \bibinfo {author} {\bibfnamefont
  {D.}~\bibnamefont {Follman}}, \bibinfo {author} {\bibfnamefont
  {P.}~\bibnamefont {Heu}}, \bibinfo {author} {\bibfnamefont {T.}~\bibnamefont
  {Corbitt}},\ and\ \bibinfo {author} {\bibfnamefont {N.}~\bibnamefont
  {Mavalvala}},\ }\href {https://doi.org/10.1038/s41567-020-0877-x} {\bibfield
  {journal} {\bibinfo  {journal} {Nature Physics}\ }\textbf {\bibinfo {volume}
  {16}},\ \bibinfo {pages} {784} (\bibinfo {year} {2020})}\BibitemShut
  {NoStop}%
\bibitem [{\citenamefont {Aasi}\ \emph {et~al.}(2013)\citenamefont {Aasi},
  \citenamefont {Abadie}, \citenamefont {Abbott}, \citenamefont {Abbott},
  \citenamefont {Abbott},\ and\ \citenamefont {\etal}}]{Aasi2013}%
  \BibitemOpen
  \bibfield  {author} {\bibinfo {author} {\bibfnamefont {J.}~\bibnamefont
  {Aasi}}, \bibinfo {author} {\bibfnamefont {J.}~\bibnamefont {Abadie}},
  \bibinfo {author} {\bibfnamefont {B.~P.}\ \bibnamefont {Abbott}}, \bibinfo
  {author} {\bibfnamefont {R.}~\bibnamefont {Abbott}}, \bibinfo {author}
  {\bibfnamefont {T.~D.}\ \bibnamefont {Abbott}},\ and\ \bibinfo {author}
  {\bibnamefont {\etal}},\ }\href {https://doi.org/10.1038/nphoton.2013.177}
  {\bibfield  {journal} {\bibinfo  {journal} {Nature Photonics}\ }\textbf
  {\bibinfo {volume} {7}},\ \bibinfo {pages} {613} (\bibinfo {year}
  {2013})}\BibitemShut {NoStop}%
\bibitem [{\citenamefont {Safavi-Naeini}\ \emph {et~al.}(2013)\citenamefont
  {Safavi-Naeini}, \citenamefont {Gr{\"o}blacher}, \citenamefont {Hill},
  \citenamefont {Chan}, \citenamefont {Aspelmeyer},\ and\ \citenamefont
  {Painter}}]{Safavi-Naeini2013}%
  \BibitemOpen
  \bibfield  {author} {\bibinfo {author} {\bibfnamefont {A.~H.}\ \bibnamefont
  {Safavi-Naeini}}, \bibinfo {author} {\bibfnamefont {S.}~\bibnamefont
  {Gr{\"o}blacher}}, \bibinfo {author} {\bibfnamefont {J.~T.}\ \bibnamefont
  {Hill}}, \bibinfo {author} {\bibfnamefont {J.}~\bibnamefont {Chan}}, \bibinfo
  {author} {\bibfnamefont {M.}~\bibnamefont {Aspelmeyer}},\ and\ \bibinfo
  {author} {\bibfnamefont {O.}~\bibnamefont {Painter}},\ }\href
  {https://doi.org/10.1038/nature12307} {\bibfield  {journal} {\bibinfo
  {journal} {Nature}\ }\textbf {\bibinfo {volume} {500}},\ \bibinfo {pages}
  {185} (\bibinfo {year} {2013})}\BibitemShut {NoStop}%
\bibitem [{\citenamefont {Cox}\ \emph {et~al.}(2016)\citenamefont {Cox},
  \citenamefont {Greve}, \citenamefont {Weiner},\ and\ \citenamefont
  {Thompson}}]{PhysRevLett.116.093602}%
  \BibitemOpen
  \bibfield  {author} {\bibinfo {author} {\bibfnamefont {K.~C.}\ \bibnamefont
  {Cox}}, \bibinfo {author} {\bibfnamefont {G.~P.}\ \bibnamefont {Greve}},
  \bibinfo {author} {\bibfnamefont {J.~M.}\ \bibnamefont {Weiner}},\ and\
  \bibinfo {author} {\bibfnamefont {J.~K.}\ \bibnamefont {Thompson}},\ }\href
  {https://doi.org/10.1103/PhysRevLett.116.093602} {\bibfield  {journal}
  {\bibinfo  {journal} {Phys. Rev. Lett.}\ }\textbf {\bibinfo {volume} {116}},\
  \bibinfo {pages} {093602} (\bibinfo {year} {2016})}\BibitemShut {NoStop}%
\bibitem [{\citenamefont {Andersen}\ \emph {et~al.}(2016)\citenamefont
  {Andersen}, \citenamefont {Gehring}, \citenamefont {Marquardt},\ and\
  \citenamefont {Leuchs}}]{Andersen_2016}%
  \BibitemOpen
  \bibfield  {author} {\bibinfo {author} {\bibfnamefont {U.~L.}\ \bibnamefont
  {Andersen}}, \bibinfo {author} {\bibfnamefont {T.}~\bibnamefont {Gehring}},
  \bibinfo {author} {\bibfnamefont {C.}~\bibnamefont {Marquardt}},\ and\
  \bibinfo {author} {\bibfnamefont {G.}~\bibnamefont {Leuchs}},\ }\href
  {https://doi.org/10.1088/0031-8949/91/5/053001} {\bibfield  {journal}
  {\bibinfo  {journal} {Physica Scripta}\ }\textbf {\bibinfo {volume} {91}},\
  \bibinfo {pages} {053001} (\bibinfo {year} {2016})}\BibitemShut {NoStop}%
\bibitem [{\citenamefont {Yap}\ \emph {et~al.}(2020)\citenamefont {Yap},
  \citenamefont {Cripe}, \citenamefont {Mansell}, \citenamefont {McRae},
  \citenamefont {Ward}, \citenamefont {Slagmolen}, \citenamefont {Heu},
  \citenamefont {Follman}, \citenamefont {Cole}, \citenamefont {Corbitt},\ and\
  \citenamefont {McClelland}}]{squeezing3}%
  \BibitemOpen
  \bibfield  {author} {\bibinfo {author} {\bibfnamefont {M.~J.}\ \bibnamefont
  {Yap}}, \bibinfo {author} {\bibfnamefont {J.}~\bibnamefont {Cripe}}, \bibinfo
  {author} {\bibfnamefont {G.~L.}\ \bibnamefont {Mansell}}, \bibinfo {author}
  {\bibfnamefont {T.~G.}\ \bibnamefont {McRae}}, \bibinfo {author}
  {\bibfnamefont {R.~L.}\ \bibnamefont {Ward}}, \bibinfo {author}
  {\bibfnamefont {B.~J.~J.}\ \bibnamefont {Slagmolen}}, \bibinfo {author}
  {\bibfnamefont {P.}~\bibnamefont {Heu}}, \bibinfo {author} {\bibfnamefont
  {D.}~\bibnamefont {Follman}}, \bibinfo {author} {\bibfnamefont {G.~D.}\
  \bibnamefont {Cole}}, \bibinfo {author} {\bibfnamefont {T.}~\bibnamefont
  {Corbitt}},\ and\ \bibinfo {author} {\bibfnamefont {D.~E.}\ \bibnamefont
  {McClelland}},\ }\href {https://doi.org/10.1038/s41566-019-0527-y} {\bibfield
   {journal} {\bibinfo  {journal} {Nature Photonics}\ }\textbf {\bibinfo
  {volume} {14}},\ \bibinfo {pages} {19} (\bibinfo {year} {2020})}\BibitemShut
  {NoStop}%
\bibitem [{\citenamefont {Walls}(1983)}]{squeezing4}%
  \BibitemOpen
  \bibfield  {author} {\bibinfo {author} {\bibfnamefont {D.~F.}\ \bibnamefont
  {Walls}},\ }\href {https://doi.org/10.1038/306141a0} {\bibfield  {journal}
  {\bibinfo  {journal} {Nature}\ }\textbf {\bibinfo {volume} {306}},\ \bibinfo
  {pages} {141} (\bibinfo {year} {1983})}\BibitemShut {NoStop}%
\bibitem [{\citenamefont {Breitenbach}\ \emph {et~al.}(1997)\citenamefont
  {Breitenbach}, \citenamefont {Schiller},\ and\ \citenamefont
  {Mlynek}}]{squeezing5}%
  \BibitemOpen
  \bibfield  {author} {\bibinfo {author} {\bibfnamefont {G.}~\bibnamefont
  {Breitenbach}}, \bibinfo {author} {\bibfnamefont {S.}~\bibnamefont
  {Schiller}},\ and\ \bibinfo {author} {\bibfnamefont {J.}~\bibnamefont
  {Mlynek}},\ }\href {https://doi.org/10.1038/387471a0} {\bibfield  {journal}
  {\bibinfo  {journal} {Nature}\ }\textbf {\bibinfo {volume} {387}},\ \bibinfo
  {pages} {471} (\bibinfo {year} {1997})}\BibitemShut {NoStop}%
\bibitem [{\citenamefont {Zhang}\ \emph {et~al.}(2021)\citenamefont {Zhang},
  \citenamefont {Menotti}, \citenamefont {Tan}, \citenamefont {Vaidya},
  \citenamefont {Mahler}, \citenamefont {Helt}, \citenamefont {Zatti},
  \citenamefont {Liscidini}, \citenamefont {Morrison},\ and\ \citenamefont
  {Vernon}}]{Zhang2021}%
  \BibitemOpen
  \bibfield  {author} {\bibinfo {author} {\bibfnamefont {Y.}~\bibnamefont
  {Zhang}}, \bibinfo {author} {\bibfnamefont {M.}~\bibnamefont {Menotti}},
  \bibinfo {author} {\bibfnamefont {K.}~\bibnamefont {Tan}}, \bibinfo {author}
  {\bibfnamefont {V.~D.}\ \bibnamefont {Vaidya}}, \bibinfo {author}
  {\bibfnamefont {D.~H.}\ \bibnamefont {Mahler}}, \bibinfo {author}
  {\bibfnamefont {L.~G.}\ \bibnamefont {Helt}}, \bibinfo {author}
  {\bibfnamefont {L.}~\bibnamefont {Zatti}}, \bibinfo {author} {\bibfnamefont
  {M.}~\bibnamefont {Liscidini}}, \bibinfo {author} {\bibfnamefont
  {B.}~\bibnamefont {Morrison}},\ and\ \bibinfo {author} {\bibfnamefont
  {Z.}~\bibnamefont {Vernon}},\ }\href
  {https://doi.org/10.1038/s41467-021-22540-2} {\bibfield  {journal} {\bibinfo
  {journal} {Nature Communications}\ }\textbf {\bibinfo {volume} {12}},\
  \bibinfo {pages} {2233} (\bibinfo {year} {2021})}\BibitemShut {NoStop}%
\bibitem [{\citenamefont {Vernon}\ \emph {et~al.}(2019)\citenamefont {Vernon},
  \citenamefont {Quesada}, \citenamefont {Liscidini}, \citenamefont {Morrison},
  \citenamefont {Menotti}, \citenamefont {Tan},\ and\ \citenamefont
  {Sipe}}]{10.1103/PhysRevApplied.12.064024}%
  \BibitemOpen
  \bibfield  {author} {\bibinfo {author} {\bibfnamefont {Z.}~\bibnamefont
  {Vernon}}, \bibinfo {author} {\bibfnamefont {N.}~\bibnamefont {Quesada}},
  \bibinfo {author} {\bibfnamefont {M.}~\bibnamefont {Liscidini}}, \bibinfo
  {author} {\bibfnamefont {B.}~\bibnamefont {Morrison}}, \bibinfo {author}
  {\bibfnamefont {M.}~\bibnamefont {Menotti}}, \bibinfo {author} {\bibfnamefont
  {K.}~\bibnamefont {Tan}},\ and\ \bibinfo {author} {\bibfnamefont {J.~E.}\
  \bibnamefont {Sipe}},\ }\href
  {https://doi.org/10.1103/PhysRevApplied.12.064024} {\bibfield  {journal}
  {\bibinfo  {journal} {Physical Review Applied}\ }\textbf {\bibinfo {volume}
  {12}},\ \bibinfo {pages} {064024} (\bibinfo {year} {2019})}\BibitemShut
  {NoStop}%
\bibitem [{\citenamefont {Kimble}\ \emph {et~al.}(2001)\citenamefont {Kimble},
  \citenamefont {Levin}, \citenamefont {Matsko}, \citenamefont {Thorne},\ and\
  \citenamefont {Vyatchanin}}]{PhysRevD.65.022002}%
  \BibitemOpen
  \bibfield  {author} {\bibinfo {author} {\bibfnamefont {H.~J.}\ \bibnamefont
  {Kimble}}, \bibinfo {author} {\bibfnamefont {Y.}~\bibnamefont {Levin}},
  \bibinfo {author} {\bibfnamefont {A.~B.}\ \bibnamefont {Matsko}}, \bibinfo
  {author} {\bibfnamefont {K.~S.}\ \bibnamefont {Thorne}},\ and\ \bibinfo
  {author} {\bibfnamefont {S.~P.}\ \bibnamefont {Vyatchanin}},\ }\href
  {https://doi.org/10.1103/PhysRevD.65.022002} {\bibfield  {journal} {\bibinfo
  {journal} {Phys. Rev. D}\ }\textbf {\bibinfo {volume} {65}},\ \bibinfo
  {pages} {022002} (\bibinfo {year} {2001})}\BibitemShut {NoStop}%
\bibitem [{\citenamefont {Kwee}\ \emph {et~al.}(2014)\citenamefont {Kwee},
  \citenamefont {Miller}, \citenamefont {Isogai}, \citenamefont {Barsotti},\
  and\ \citenamefont {Evans}}]{PhysRevD.90.062006}%
  \BibitemOpen
  \bibfield  {author} {\bibinfo {author} {\bibfnamefont {P.}~\bibnamefont
  {Kwee}}, \bibinfo {author} {\bibfnamefont {J.}~\bibnamefont {Miller}},
  \bibinfo {author} {\bibfnamefont {T.}~\bibnamefont {Isogai}}, \bibinfo
  {author} {\bibfnamefont {L.}~\bibnamefont {Barsotti}},\ and\ \bibinfo
  {author} {\bibfnamefont {M.}~\bibnamefont {Evans}},\ }\href
  {https://doi.org/10.1103/PhysRevD.90.062006} {\bibfield  {journal} {\bibinfo
  {journal} {Phys. Rev. D}\ }\textbf {\bibinfo {volume} {90}},\ \bibinfo
  {pages} {062006} (\bibinfo {year} {2014})}\BibitemShut {NoStop}%
\bibitem [{\citenamefont {Dutt}\ \emph {et~al.}(2016)\citenamefont {Dutt},
  \citenamefont {Miller}, \citenamefont {Luke}, \citenamefont {Cardenas},
  \citenamefont {Gaeta}, \citenamefont {Nussenzveig},\ and\ \citenamefont
  {Lipson}}]{Dutt:16}%
  \BibitemOpen
  \bibfield  {author} {\bibinfo {author} {\bibfnamefont {A.}~\bibnamefont
  {Dutt}}, \bibinfo {author} {\bibfnamefont {S.}~\bibnamefont {Miller}},
  \bibinfo {author} {\bibfnamefont {K.}~\bibnamefont {Luke}}, \bibinfo {author}
  {\bibfnamefont {J.}~\bibnamefont {Cardenas}}, \bibinfo {author}
  {\bibfnamefont {A.~L.}\ \bibnamefont {Gaeta}}, \bibinfo {author}
  {\bibfnamefont {P.}~\bibnamefont {Nussenzveig}},\ and\ \bibinfo {author}
  {\bibfnamefont {M.}~\bibnamefont {Lipson}},\ }\href
  {https://doi.org/10.1364/OL.41.000223} {\bibfield  {journal} {\bibinfo
  {journal} {Opt. Lett.}\ }\textbf {\bibinfo {volume} {41}},\ \bibinfo {pages}
  {223} (\bibinfo {year} {2016})}\BibitemShut {NoStop}%
\bibitem [{\citenamefont {Davuluri}\ and\ \citenamefont
  {Li}(2016)}]{Davuluri_2016}%
  \BibitemOpen
  \bibfield  {author} {\bibinfo {author} {\bibfnamefont {S.}~\bibnamefont
  {Davuluri}}\ and\ \bibinfo {author} {\bibfnamefont {Y.}~\bibnamefont {Li}},\
  }\href {https://doi.org/10.1088/1367-2630/18/10/103047} {\bibfield  {journal}
  {\bibinfo  {journal} {New Journal of Physics}\ }\textbf {\bibinfo {volume}
  {18}},\ \bibinfo {pages} {103047} (\bibinfo {year} {2016})}\BibitemShut
  {NoStop}%
\bibitem [{\citenamefont {Vahlbruch}\ \emph {et~al.}(2016)\citenamefont
  {Vahlbruch}, \citenamefont {Mehmet}, \citenamefont {Danzmann},\ and\
  \citenamefont {Schnabel}}]{PhysRevLett.117.110801}%
  \BibitemOpen
  \bibfield  {author} {\bibinfo {author} {\bibfnamefont {H.}~\bibnamefont
  {Vahlbruch}}, \bibinfo {author} {\bibfnamefont {M.}~\bibnamefont {Mehmet}},
  \bibinfo {author} {\bibfnamefont {K.}~\bibnamefont {Danzmann}},\ and\
  \bibinfo {author} {\bibfnamefont {R.}~\bibnamefont {Schnabel}},\ }\href
  {https://doi.org/10.1103/PhysRevLett.117.110801} {\bibfield  {journal}
  {\bibinfo  {journal} {Phys. Rev. Lett.}\ }\textbf {\bibinfo {volume} {117}},\
  \bibinfo {pages} {110801} (\bibinfo {year} {2016})}\BibitemShut {NoStop}%
\bibitem [{\citenamefont {Burgwal}\ \emph {et~al.}(2020)\citenamefont
  {Burgwal}, \citenamefont {del Pino},\ and\ \citenamefont
  {Verhagen}}]{Burgwal_2020}%
  \BibitemOpen
  \bibfield  {author} {\bibinfo {author} {\bibfnamefont {R.}~\bibnamefont
  {Burgwal}}, \bibinfo {author} {\bibfnamefont {J.}~\bibnamefont {del Pino}},\
  and\ \bibinfo {author} {\bibfnamefont {E.}~\bibnamefont {Verhagen}},\ }\href
  {https://doi.org/10.1088/1367-2630/abc1c8} {\bibfield  {journal} {\bibinfo
  {journal} {New Journal of Physics}\ }\textbf {\bibinfo {volume} {22}},\
  \bibinfo {pages} {113006} (\bibinfo {year} {2020})}\BibitemShut {NoStop}%
\bibitem [{\citenamefont {Law}(1995)}]{PhysRevA.51.2537}%
  \BibitemOpen
  \bibfield  {author} {\bibinfo {author} {\bibfnamefont {C.~K.}\ \bibnamefont
  {Law}},\ }\href {https://doi.org/10.1103/PhysRevA.51.2537} {\bibfield
  {journal} {\bibinfo  {journal} {Phys. Rev. A}\ }\textbf {\bibinfo {volume}
  {51}},\ \bibinfo {pages} {2537} (\bibinfo {year} {1995})}\BibitemShut
  {NoStop}%
\bibitem [{\citenamefont {Ludwig}\ \emph {et~al.}(2012)\citenamefont {Ludwig},
  \citenamefont {Safavi-Naeini}, \citenamefont {Painter},\ and\ \citenamefont
  {Marquardt}}]{PhysRevLett.109.063601}%
  \BibitemOpen
  \bibfield  {author} {\bibinfo {author} {\bibfnamefont {M.}~\bibnamefont
  {Ludwig}}, \bibinfo {author} {\bibfnamefont {A.~H.}\ \bibnamefont
  {Safavi-Naeini}}, \bibinfo {author} {\bibfnamefont {O.}~\bibnamefont
  {Painter}},\ and\ \bibinfo {author} {\bibfnamefont {F.}~\bibnamefont
  {Marquardt}},\ }\href {https://doi.org/10.1103/PhysRevLett.109.063601}
  {\bibfield  {journal} {\bibinfo  {journal} {Phys. Rev. Lett.}\ }\textbf
  {\bibinfo {volume} {109}},\ \bibinfo {pages} {063601} (\bibinfo {year}
  {2012})}\BibitemShut {NoStop}%
\bibitem [{\citenamefont {Grudinin}\ \emph {et~al.}(2010)\citenamefont
  {Grudinin}, \citenamefont {Lee}, \citenamefont {Painter},\ and\ \citenamefont
  {Vahala}}]{PhysRevLett.104.083901}%
  \BibitemOpen
  \bibfield  {author} {\bibinfo {author} {\bibfnamefont {I.~S.}\ \bibnamefont
  {Grudinin}}, \bibinfo {author} {\bibfnamefont {H.}~\bibnamefont {Lee}},
  \bibinfo {author} {\bibfnamefont {O.}~\bibnamefont {Painter}},\ and\ \bibinfo
  {author} {\bibfnamefont {K.~J.}\ \bibnamefont {Vahala}},\ }\href
  {https://doi.org/10.1103/PhysRevLett.104.083901} {\bibfield  {journal}
  {\bibinfo  {journal} {Phys. Rev. Lett.}\ }\textbf {\bibinfo {volume} {104}},\
  \bibinfo {pages} {083901} (\bibinfo {year} {2010})}\BibitemShut {NoStop}%
\bibitem [{\citenamefont {Dorsel}\ \emph {et~al.}(1983)\citenamefont {Dorsel},
  \citenamefont {McCullen}, \citenamefont {Meystre}, \citenamefont {Vignes},\
  and\ \citenamefont {Walther}}]{PhysRevLett.51.1550}%
  \BibitemOpen
  \bibfield  {author} {\bibinfo {author} {\bibfnamefont {A.}~\bibnamefont
  {Dorsel}}, \bibinfo {author} {\bibfnamefont {J.~D.}\ \bibnamefont
  {McCullen}}, \bibinfo {author} {\bibfnamefont {P.}~\bibnamefont {Meystre}},
  \bibinfo {author} {\bibfnamefont {E.}~\bibnamefont {Vignes}},\ and\ \bibinfo
  {author} {\bibfnamefont {H.}~\bibnamefont {Walther}},\ }\href
  {https://doi.org/10.1103/PhysRevLett.51.1550} {\bibfield  {journal} {\bibinfo
   {journal} {Phys. Rev. Lett.}\ }\textbf {\bibinfo {volume} {51}},\ \bibinfo
  {pages} {1550} (\bibinfo {year} {1983})}\BibitemShut {NoStop}%
\bibitem [{\citenamefont {Davuluri}(2021)}]{davuluri2017quantum}%
  \BibitemOpen
  \bibfield  {author} {\bibinfo {author} {\bibfnamefont {S.}~\bibnamefont
  {Davuluri}},\ }\href {https://doi.org/10.1364/OL.412822} {\bibfield
  {journal} {\bibinfo  {journal} {Opt. Lett.}\ }\textbf {\bibinfo {volume}
  {46}},\ \bibinfo {pages} {904} (\bibinfo {year} {2021})}\BibitemShut
  {NoStop}%
\bibitem [{\citenamefont {Giovannetti}\ and\ \citenamefont
  {Vitali}(2001)}]{PhysRevA.63.023812}%
  \BibitemOpen
  \bibfield  {author} {\bibinfo {author} {\bibfnamefont {V.}~\bibnamefont
  {Giovannetti}}\ and\ \bibinfo {author} {\bibfnamefont {D.}~\bibnamefont
  {Vitali}},\ }\href {https://doi.org/10.1103/PhysRevA.63.023812} {\bibfield
  {journal} {\bibinfo  {journal} {Phys. Rev. A}\ }\textbf {\bibinfo {volume}
  {63}},\ \bibinfo {pages} {023812} (\bibinfo {year} {2001})}\BibitemShut
  {NoStop}%
\bibitem [{\citenamefont
  {Lvovsky}(2015)}]{doi:https://doi.org/10.1002/9781119009719.ch5}%
  \BibitemOpen
  \bibfield  {author} {\bibinfo {author} {\bibfnamefont {A.~I.}\ \bibnamefont
  {Lvovsky}},\ }in\ \href
  {https://doi.org/https://doi.org/10.1002/9781119009719.ch5} {\emph {\bibinfo
  {booktitle} {Photonics}}}\ (\bibinfo  {publisher} {John Wiley and Sons,
  Ltd},\ \bibinfo {year} {2015})\ Chap.~\bibinfo {chapter} {5}, pp.\ \bibinfo
  {pages} {121--163}\BibitemShut {NoStop}%
\bibitem [{\citenamefont {Cernansky}\ and\ \citenamefont
  {Politi}(2020)}]{doi:10.1063/5.0024341}%
  \BibitemOpen
  \bibfield  {author} {\bibinfo {author} {\bibfnamefont {R.}~\bibnamefont
  {Cernansky}}\ and\ \bibinfo {author} {\bibfnamefont {A.}~\bibnamefont
  {Politi}},\ }\href {https://doi.org/10.1063/5.0024341} {\bibfield  {journal}
  {\bibinfo  {journal} {APL Photonics}\ }\textbf {\bibinfo {volume} {5}},\
  \bibinfo {pages} {101303} (\bibinfo {year} {2020})}\BibitemShut {NoStop}%
\bibitem [{\citenamefont {Stoler}(1970)}]{PhysRevD.1.3217}%
  \BibitemOpen
  \bibfield  {author} {\bibinfo {author} {\bibfnamefont {D.}~\bibnamefont
  {Stoler}},\ }\href {https://doi.org/10.1103/PhysRevD.1.3217} {\bibfield
  {journal} {\bibinfo  {journal} {Phys. Rev. D}\ }\textbf {\bibinfo {volume}
  {1}},\ \bibinfo {pages} {3217} (\bibinfo {year} {1970})}\BibitemShut
  {NoStop}%
\bibitem [{\citenamefont {Stoler}(1971)}]{PhysRevD.4.1925}%
  \BibitemOpen
  \bibfield  {author} {\bibinfo {author} {\bibfnamefont {D.}~\bibnamefont
  {Stoler}},\ }\href {https://doi.org/10.1103/PhysRevD.4.1925} {\bibfield
  {journal} {\bibinfo  {journal} {Phys. Rev. D}\ }\textbf {\bibinfo {volume}
  {4}},\ \bibinfo {pages} {1925} (\bibinfo {year} {1971})}\BibitemShut
  {NoStop}%
\bibitem [{\citenamefont {Otterpohl}\ \emph {et~al.}(2019)\citenamefont
  {Otterpohl}, \citenamefont {Sedlmeir}, \citenamefont {Vogl}, \citenamefont
  {Dirmeier}, \citenamefont {Shafiee}, \citenamefont {Schunk}, \citenamefont
  {Strekalov}, \citenamefont {Schwefel}, \citenamefont {Gehring}, \citenamefont
  {Andersen}, \citenamefont {Leuchs},\ and\ \citenamefont
  {Marquardt}}]{Otterpohl:19}%
  \BibitemOpen
  \bibfield  {author} {\bibinfo {author} {\bibfnamefont {A.}~\bibnamefont
  {Otterpohl}}, \bibinfo {author} {\bibfnamefont {F.}~\bibnamefont {Sedlmeir}},
  \bibinfo {author} {\bibfnamefont {U.}~\bibnamefont {Vogl}}, \bibinfo {author}
  {\bibfnamefont {T.}~\bibnamefont {Dirmeier}}, \bibinfo {author}
  {\bibfnamefont {G.}~\bibnamefont {Shafiee}}, \bibinfo {author} {\bibfnamefont
  {G.}~\bibnamefont {Schunk}}, \bibinfo {author} {\bibfnamefont {D.~V.}\
  \bibnamefont {Strekalov}}, \bibinfo {author} {\bibfnamefont {H.~G.~L.}\
  \bibnamefont {Schwefel}}, \bibinfo {author} {\bibfnamefont {T.}~\bibnamefont
  {Gehring}}, \bibinfo {author} {\bibfnamefont {U.~L.}\ \bibnamefont
  {Andersen}}, \bibinfo {author} {\bibfnamefont {G.}~\bibnamefont {Leuchs}},\
  and\ \bibinfo {author} {\bibfnamefont {C.}~\bibnamefont {Marquardt}},\ }\href
  {https://doi.org/10.1364/OPTICA.6.001375} {\bibfield  {journal} {\bibinfo
  {journal} {Optica}\ }\textbf {\bibinfo {volume} {6}},\ \bibinfo {pages}
  {1375} (\bibinfo {year} {2019})}\BibitemShut {NoStop}%
\bibitem [{\citenamefont {Ast}\ \emph {et~al.}(2013)\citenamefont {Ast},
  \citenamefont {Mehmet},\ and\ \citenamefont {Schnabel}}]{Ast:13}%
  \BibitemOpen
  \bibfield  {author} {\bibinfo {author} {\bibfnamefont {S.}~\bibnamefont
  {Ast}}, \bibinfo {author} {\bibfnamefont {M.}~\bibnamefont {Mehmet}},\ and\
  \bibinfo {author} {\bibfnamefont {R.}~\bibnamefont {Schnabel}},\ }\href
  {https://doi.org/10.1364/OE.21.013572} {\bibfield  {journal} {\bibinfo
  {journal} {Opt. Express}\ }\textbf {\bibinfo {volume} {21}},\ \bibinfo
  {pages} {13572} (\bibinfo {year} {2013})}\BibitemShut {NoStop}%
\bibitem [{\citenamefont {Aoki}\ \emph {et~al.}(2006)\citenamefont {Aoki},
  \citenamefont {Takahashi},\ and\ \citenamefont {Furusawa}}]{Aoki:06}%
  \BibitemOpen
  \bibfield  {author} {\bibinfo {author} {\bibfnamefont {T.}~\bibnamefont
  {Aoki}}, \bibinfo {author} {\bibfnamefont {G.}~\bibnamefont {Takahashi}},\
  and\ \bibinfo {author} {\bibfnamefont {A.}~\bibnamefont {Furusawa}},\ }\href
  {http://www.osapublishing.org/oe/} {\bibfield  {journal} {\bibinfo  {journal}
  {Opt. Express}\ }\textbf {\bibinfo {volume} {14}},\ \bibinfo {pages} {6930}
  (\bibinfo {year} {2006})}\BibitemShut {NoStop}%
\bibitem [{\citenamefont {Zhao}\ \emph {et~al.}(2020)\citenamefont {Zhao},
  \citenamefont {Okawachi}, \citenamefont {Jang}, \citenamefont {Ji},
  \citenamefont {Lipson},\ and\ \citenamefont
  {Gaeta}}]{PhysRevLett.124.193601}%
  \BibitemOpen
  \bibfield  {author} {\bibinfo {author} {\bibfnamefont {Y.}~\bibnamefont
  {Zhao}}, \bibinfo {author} {\bibfnamefont {Y.}~\bibnamefont {Okawachi}},
  \bibinfo {author} {\bibfnamefont {J.~K.}\ \bibnamefont {Jang}}, \bibinfo
  {author} {\bibfnamefont {X.}~\bibnamefont {Ji}}, \bibinfo {author}
  {\bibfnamefont {M.}~\bibnamefont {Lipson}},\ and\ \bibinfo {author}
  {\bibfnamefont {A.~L.}\ \bibnamefont {Gaeta}},\ }\href
  {https://doi.org/10.1103/PhysRevLett.124.193601} {\bibfield  {journal}
  {\bibinfo  {journal} {Phys. Rev. Lett.}\ }\textbf {\bibinfo {volume} {124}},\
  \bibinfo {pages} {193601} (\bibinfo {year} {2020})}\BibitemShut {NoStop}%
\bibitem [{\citenamefont {Braginsky}\ and\ \citenamefont
  {Khalili}(1996)}]{RevModPhys.68.1}%
  \BibitemOpen
  \bibfield  {author} {\bibinfo {author} {\bibfnamefont {V.~B.}\ \bibnamefont
  {Braginsky}}\ and\ \bibinfo {author} {\bibfnamefont {F.~Y.}\ \bibnamefont
  {Khalili}},\ }\href {https://doi.org/10.1103/RevModPhys.68.1} {\bibfield
  {journal} {\bibinfo  {journal} {Rev. Mod. Phys.}\ }\textbf {\bibinfo {volume}
  {68}},\ \bibinfo {pages} {1} (\bibinfo {year} {1996})}\BibitemShut {NoStop}%
\bibitem [{\citenamefont {Tsang}\ and\ \citenamefont
  {Caves}(2012)}]{PhysRevX.2.031016}%
  \BibitemOpen
  \bibfield  {author} {\bibinfo {author} {\bibfnamefont {M.}~\bibnamefont
  {Tsang}}\ and\ \bibinfo {author} {\bibfnamefont {C.~M.}\ \bibnamefont
  {Caves}},\ }\href {https://doi.org/10.1103/PhysRevX.2.031016} {\bibfield
  {journal} {\bibinfo  {journal} {Phys. Rev. X}\ }\textbf {\bibinfo {volume}
  {2}},\ \bibinfo {pages} {031016} (\bibinfo {year} {2012})}\BibitemShut
  {NoStop}%
\bibitem [{\citenamefont {Li}\ \emph {et~al.}(2018)\citenamefont {Li},
  \citenamefont {Davuluri},\ and\ \citenamefont {Li}}]{Li2018}%
  \BibitemOpen
  \bibfield  {author} {\bibinfo {author} {\bibfnamefont {K.}~\bibnamefont
  {Li}}, \bibinfo {author} {\bibfnamefont {S.}~\bibnamefont {Davuluri}},\ and\
  \bibinfo {author} {\bibfnamefont {Y.}~\bibnamefont {Li}},\ }\href
  {https://doi.org/10.1007/s11433-018-9189-6} {\bibfield  {journal} {\bibinfo
  {journal} {Science China Physics, Mechanics {\&} Astronomy}\ }\textbf
  {\bibinfo {volume} {61}},\ \bibinfo {pages} {90311} (\bibinfo {year}
  {2018})}\BibitemShut {NoStop}%
\bibitem [{\citenamefont {Kampel}\ \emph {et~al.}(2017)\citenamefont {Kampel},
  \citenamefont {Peterson}, \citenamefont {Fischer}, \citenamefont {Yu},
  \citenamefont {Cicak}, \citenamefont {Simmonds}, \citenamefont {Lehnert},\
  and\ \citenamefont {Regal}}]{PhysRevX.7.021008}%
  \BibitemOpen
  \bibfield  {author} {\bibinfo {author} {\bibfnamefont {N.~S.}\ \bibnamefont
  {Kampel}}, \bibinfo {author} {\bibfnamefont {R.~W.}\ \bibnamefont
  {Peterson}}, \bibinfo {author} {\bibfnamefont {R.}~\bibnamefont {Fischer}},
  \bibinfo {author} {\bibfnamefont {P.-L.}\ \bibnamefont {Yu}}, \bibinfo
  {author} {\bibfnamefont {K.}~\bibnamefont {Cicak}}, \bibinfo {author}
  {\bibfnamefont {R.~W.}\ \bibnamefont {Simmonds}}, \bibinfo {author}
  {\bibfnamefont {K.~W.}\ \bibnamefont {Lehnert}},\ and\ \bibinfo {author}
  {\bibfnamefont {C.~A.}\ \bibnamefont {Regal}},\ }\href
  {https://doi.org/10.1103/PhysRevX.7.021008} {\bibfield  {journal} {\bibinfo
  {journal} {Phys. Rev. X}\ }\textbf {\bibinfo {volume} {7}},\ \bibinfo {pages}
  {021008} (\bibinfo {year} {2017})}\BibitemShut {NoStop}%
\bibitem [{\citenamefont {de~Lépinay}\ \emph {et~al.}(2021)\citenamefont
  {de~Lépinay}, \citenamefont {Ockeloen-Korppi}, \citenamefont {Woolley},\
  and\ \citenamefont {Sillanpää}}]{doi:10.1126/science.abf5389}%
  \BibitemOpen
  \bibfield  {author} {\bibinfo {author} {\bibfnamefont {L.~M.}\ \bibnamefont
  {de~Lépinay}}, \bibinfo {author} {\bibfnamefont {C.~F.}\ \bibnamefont
  {Ockeloen-Korppi}}, \bibinfo {author} {\bibfnamefont {M.~J.}\ \bibnamefont
  {Woolley}},\ and\ \bibinfo {author} {\bibfnamefont {M.~A.}\ \bibnamefont
  {Sillanpää}},\ }\href {https://doi.org/10.1126/science.abf5389} {\bibfield
  {journal} {\bibinfo  {journal} {Science}\ }\textbf {\bibinfo {volume}
  {372}},\ \bibinfo {pages} {625} (\bibinfo {year} {2021})}\BibitemShut
  {NoStop}%
\bibitem [{\citenamefont {Buonanno}\ and\ \citenamefont
  {Chen}(2001)}]{PhysRevD.64.042006}%
  \BibitemOpen
  \bibfield  {author} {\bibinfo {author} {\bibfnamefont {A.}~\bibnamefont
  {Buonanno}}\ and\ \bibinfo {author} {\bibfnamefont {Y.}~\bibnamefont
  {Chen}},\ }\href {https://doi.org/10.1103/PhysRevD.64.042006} {\bibfield
  {journal} {\bibinfo  {journal} {Phys. Rev. D}\ }\textbf {\bibinfo {volume}
  {64}},\ \bibinfo {pages} {042006} (\bibinfo {year} {2001})}\BibitemShut
  {NoStop}%
\bibitem [{\citenamefont {Buonanno}\ and\ \citenamefont
  {Chen}(2003)}]{PhysRevD.67.062002}%
  \BibitemOpen
  \bibfield  {author} {\bibinfo {author} {\bibfnamefont {A.}~\bibnamefont
  {Buonanno}}\ and\ \bibinfo {author} {\bibfnamefont {Y.}~\bibnamefont
  {Chen}},\ }\href {https://doi.org/10.1103/PhysRevD.67.062002} {\bibfield
  {journal} {\bibinfo  {journal} {Phys. Rev. D}\ }\textbf {\bibinfo {volume}
  {67}},\ \bibinfo {pages} {062002} (\bibinfo {year} {2003})}\BibitemShut
  {NoStop}%
\bibitem [{\citenamefont {Rehbein}\ \emph {et~al.}(2007)\citenamefont
  {Rehbein}, \citenamefont {M\"uller-Ebhardt}, \citenamefont {Somiya},
  \citenamefont {Li}, \citenamefont {Schnabel}, \citenamefont {Danzmann},\ and\
  \citenamefont {Chen}}]{item_150019}%
  \BibitemOpen
  \bibfield  {author} {\bibinfo {author} {\bibfnamefont {H.}~\bibnamefont
  {Rehbein}}, \bibinfo {author} {\bibfnamefont {H.}~\bibnamefont
  {M\"uller-Ebhardt}}, \bibinfo {author} {\bibfnamefont {K.}~\bibnamefont
  {Somiya}}, \bibinfo {author} {\bibfnamefont {C.}~\bibnamefont {Li}}, \bibinfo
  {author} {\bibfnamefont {R.}~\bibnamefont {Schnabel}}, \bibinfo {author}
  {\bibfnamefont {K.}~\bibnamefont {Danzmann}},\ and\ \bibinfo {author}
  {\bibfnamefont {Y.}~\bibnamefont {Chen}},\ }\href
  {https://doi.org/10.1103/PhysRevD.76.062002} {\bibfield  {journal} {\bibinfo
  {journal} {Phys. Rev. D}\ }\textbf {\bibinfo {volume} {76}},\ \bibinfo
  {pages} {062002} (\bibinfo {year} {2007})}\BibitemShut {NoStop}%
\bibitem [{\citenamefont {Davuluri}\ and\ \citenamefont
  {Li}(2022)}]{davuluri2022light}%
  \BibitemOpen
  \bibfield  {author} {\bibinfo {author} {\bibfnamefont {S.}~\bibnamefont
  {Davuluri}}\ and\ \bibinfo {author} {\bibfnamefont {Y.}~\bibnamefont {Li}},\
  }\href {https://link.aps.org/doi/10.1103/PhysRevD.4.1925} {\bibfield
  {journal} {\bibinfo  {journal} {arXiv preprint arXiv:2202.06030}\ } (\bibinfo
  {year} {2022})}\BibitemShut {NoStop}%
\end{thebibliography}%
\end{document}